\documentclass[apj,iop]{emulateapj}
\usepackage{graphicx}   
\usepackage{amsmath}

\usepackage{color}
\usepackage[dvipsnames]{xcolor}

\newcommand{\dM}{\ensuremath{\Delta M}}
\newcommand{\sigphot}{\ensuremath{\sigma_{\rm phot}}}
\newcommand{\DL}{\ensuremath{\mathcal{D}_L (z; \Omega_M, w)}}
\newcommand{\scriptM}{\ensuremath{\mathcal{M}}}

\begin{document}
\title{Incorporating Astrophysical Systematics into a Generalized Likelihood for Cosmology with Type Ia Supernovae}

\author{
Kara~A.~Ponder*, W.~Michael~Wood-Vasey, Andrew~R.~Zentner
}

\affil{
Pittsburgh Particle Physics, Astrophysics, and Cosmology Center (PITT PACC). \\
Physics and Astronomy Department, University of Pittsburgh, 
Pittsburgh PA, 15260, USA
}

\email{* Email: kap146@pitt.edu}



\begin{abstract}

Traditional cosmological inference using Type Ia supernovae (SNeIa) have used stretch- and color-corrected fits of SN Ia light curves and assumed a resulting fiducial mean and symmetric intrinsic dispersion for the resulting relative luminosity. 
As systematics become the main contributors to the error budget,
it has become imperative to expand supernova cosmology analyses to include a more general likelihood to model systematics to remove biases with losses in precision. 
To illustrate an example likelihood analysis,
 we use a simple model of two populations with a relative luminosity shift, 
 independent intrinsic dispersions, and linear redshift evolution of the relative fraction of each population. 
Treating observationally viable two-population mock data using a one-population model results in an inferred dark energy equation of state parameter $w$ that is biased by roughly 2 times its statistical error for a sample of N $ \gtrsim$ 2500 SNeIa. 
Modeling the two-population data with a two-population model removes this bias at a cost of an approximately $\sim20\%$ increase in the statistical constraint on $w$. 
These significant biases can be realized even if the support for two underlying SNeIa populations, in the form of model selection criteria, is inconclusive. 
With the current observationally-estimated difference in the two proposed populations, a sample of N $ \gtrsim$ 10,000 SNeIa is necessary to yield conclusive evidence of two populations.

\end{abstract}


\section{Introduction}\label{sec:introduction}

Type Ia supernovae (SNeIa) are excellent standardizable candles that enabled the discovery of the expansion of the universe in the late 1990s by \cite{Riess98} and \cite{Perlmutter99}. 
Originally, SNeIa were used as standard candles from empirical evidence with a scatter of only $\sim$0.3~magnitudes  \citep{Baade38, Kowal68}. 
As data sets grew, patterns appeared in the light curves yielding the brighter-slower \citep{Phillips93} and brighter-bluer  \citep{Riess96, Tripp98} relationships, which
 standardized the supernovae further by reducing their scatter down to $\sim 0.15$ magnitudes. 

The goal of this paper is to create a framework to 
properly model effects that change the {\em distribution} of expected
SN~Ia apparent brightness at each redshift.
If unmodeled, these effects lead to systematic biases in cosmological inference.
We propose using general and flexible likelihood functions that have the ability to handle insufficiently modeled systematics.
As an example, we simulate a simplistic toy model of two SN~Ia populations 
with a small relative shift in absolute magnitude.
The relative rate of these two populations changes linearly with redshift.
We examine the systematic errors in cosmological parameters caused by incorrectly fitting multiple 
populations with a single Gaussian model and show that these errors can be eliminated by using a 
multiple population model to fit the SN~Ia magnitude--redshift relation. 
In this paper, we focus on this toy model to demonstrate the validity of this framework.  
The consideration of more complex multiple-population models or other astrophysical or observational effects that lead to shifting magnitude distributions with redshift will be considered in subsequent papers. 

Though the two population model is intended as an example, there are several motivators for multiple populations of SNeIa. 
For instance,
after adjusting the light curves with these observed relationships, there is still an unaccounted for feature in the corrected brightness residual with respect to the distance-redshift relationship (Hubble residual) that appears to be correlated with host galaxy properties. 
In the last five years, there have been myriad studies \citep{Kelly10, Sullivan10, Lampeitl10, Gupta11, Johansson13, Childress13a, Rigault13, Rigault15, Kelly15} comparing host galaxy mass, metallicity, and/or star formation rate to residuals in the Hubble diagram. 
 \cite{Rigault13}  examined the relationship between global and local star formation rates through H-alpha 
and found that SNeIa in locally passive environments were brighter than those in locally star forming environments.  
\cite{Rigault15} and \citet{Kelly15} used {\it GALEX} ultraviolet observations 
and confirmed this correlation between Hubble residual and local star-formation rate.
 
It is possible that the host galaxy correlations are caused by something more fundamental such as the nature of the progenitor. 
Though the evidence for host galaxy correlations may be controversial \citep{Jones15},
there is increasing evidence that there are two different progenitor channels that could create a multiple population effect \citep{Greggio05, Cao15, Olling15}.

The most recent analysis of SNeIa for cosmology comes from \cite{Betoule14} with the Joint Lightcurve Analysis (JLA). 
They account for the observed correlation between Hubble residual and host galaxy mass by creating a step function for the absolute magnitude of each supernova based on the host galaxy mass. 
They then implicitly assume a Gaussian likelihood and fit for parameters using a $\chi^2$ method. 
We will expand this method by defining a continuous function for absolute magnitude and finding the most probable parameter regions with a generalized likelihood through Markov Chain Monte Carlo (MCMC) techniques.

Being able to identify and accurately correct for systematics is becoming more important as the number of SNeIa drastically increases with current surveys such as Dark Energy Survey (DES)\footnote{http://www.darkenergysurvey.org/}, 
Panoramic Survey Telescope and Rapid Response System \citep[Pan-STARRS][]{Scolnic14,Rest14}. 
The amount of SN~Ia data available for cosmological analyses will continue to increase into the future with surveys such as  the
Large Synoptic Survey Telescope (LSST, \cite{LSST09}), Wide-Field Infrared Survey Telescope-Astrophysics Focused Telescope Asset (WFIRST-AFTA, \cite{WFIRST15}), 
and the European Space Agency's Euclid\footnote{\url{http://sci.esa.int/euclid/}} mission on the horizon~\citep{Astier14}. 
Supernova cosmology is no longer statistically limited and is rapidly becoming systematically limited. 
Now is the time to explore different avenues for undertaking unbiased cosmological analyses 
with large data sets.

In Section~\ref{sec:NGED} we discuss non-Gaussian error distributions as modeled by multiple Gaussian populations.
Section~\ref{sec:datasets} defines how mock SN~Ia data sets are generated. 
Here, we introduce a toy model that represents a redshift evolution of the populations to probe uses of the framework.
In Section~\ref{sec:methods} we define the likelihood to be used in the MCMC and the different model selection techniques.
Section~\ref{sec:results} shows that both population and cosmological parameters are biased if multiple populations are not included in the analysis.
Though it has more model parameters, the Gaussian mixture model recovers input cosmology with only a $\sim1-3\%$ loss in precision.
We show that current and future data sets will have the cosmology biased before there are enough statistics to characterize the underlying systematic or 
to robustly require a more complicated model.
Section~\ref{sec:discussion} discusses how the models presented here relate to current cosmological analyses and presents possible astrophysical motivations for multiple populations.
In Section~\ref{sec:conclusion} we summarize our results and discuss ways to improve and expand this framework.


\section{Non-Gaussian Error Distributions}\label{sec:NGED}
The most commonly used method for cosmological parameter estimation in supernova cosmology is $\chi^2$ minimization. 
Implicit in this method is the assumption that the overall probability density function (PDF) of supernovae follows a Gaussian distribution or can be linearly corrected to do so.
With motivations such as the host galaxy correlations, complexities in the analysis from Malmquist bias, and uncertainties about dust, 
there are too many unknowns within supernova cosmology for SNeIa to be adequately described by a single point estimator in a Gaussian PDF. 
The PDF of SNeIa needs to be expanded to be able to more fully model the systematic effects underlying the observed luminosity distribution of SNeIa.

Here we will explore one possible expansion to the PDF of SNIa luminosity.
While this example is inspired by the recent discussions of correlations between
SNIa corrected luminosity and host galaxy properties, 
its use here is intended as a demonstration of the framework.
We are not arguing for any particular specific model as 
being representative of the SNIa population extant in the Universe.

\subsection{Gaussian Mixture Models}\label{sec:GMM}
Karl Pearson popularized using multiple Gaussians to describe non-Gaussian data in 1894 when he showed that two Gaussians were a better fit to crab morphologies which strengthened the claim for evolution \citep{Pearson94}.\footnote{Thanks to S. Peng Oh for this reference.}

A distribution consisting of multiple Gaussian populations with different peaks and/or dispersions is referred to as a Gaussian mixture model (GMM) and the probability density function (PDF) that describes it is

\begin{equation}
\label{eq:GMM}
p_{\rm GMM} (x) = \sum\limits_{j=1}^\mathcal{N} \frac{n_j}{\sqrt{2 \pi \sigma_j^2}} \exp \left( \frac{-(x-\lambda_j)^2}{2 \sigma_j^2} \right)
\end{equation}
where $\mathcal{N}$ is the number of populations; and for each population $j$: $n_j$ is the relative normalization $\left(\sum\limits_{j=1}^\mathcal{N} n_j = 1\right)$; $\lambda_j$ is the mean; and $\sigma_j$ is the standard deviation.

For the sake of simplicity and because it is motivated by current observational literature, in this paper we focus on a model with 
only two populations: A and B. Under this model Eq.~\ref{eq:GMM} then becomes 

\begin{eqnarray}
\label{eq:GMM_prob}
p_{\rm GMM}(x) & = & \frac{n_A}{\sqrt{2 \pi \sigma_A^2}} \exp \left( \frac{-(x - \lambda_A)^2}{2 \sigma_A^2} \right) +  \nonumber \\
               &   & \frac{n_B}{\sqrt{2 \pi \sigma_B^2}} \exp \left( \frac{-(x - \lambda_B)^2}{2 \sigma_B^2} \right) .
\end{eqnarray}

There are five parameters that need to be specified: $\lambda_A$, $\lambda_B$, $\sigma_A$, $\sigma_B$, and $n_A$ ($n_B$ is implicitly specified under the constraint that $n_A+n_B = 1$). Once the PDF has been defined,
the log-likelihood function for the two-population model, $\mathcal{L}$, is simply 

\begin{eqnarray}
\mathcal{L} = \ln L = \sum\limits_{i=1}^N \ln \Big [ & \frac{n_A}{\sqrt{2 \pi \sigma_A^2}} \exp \left(\frac{-(x_i - \lambda_A)^2}{2 \sigma_A^2} \right) + \nonumber \\
        & \frac{n_B}{\sqrt{2 \pi \sigma_B^2}} \exp \left(\frac{-(x_i - \lambda_B)^2}{2 \sigma_B^2} \right) \Big ],
\label{eq:GMM_LL_lambda}
\end{eqnarray}
where $N$ is the total number of objects included in the analysis, $x_i$ is some observed quantity per object, and ($\lambda_A$, $\sigma_A$), ($\lambda_B$, $\sigma_B$) are the model mean and standard deviation for the populations A and B. \\

In the case of SN~Ia cosmology, $x_i$ is the observed width-color-corrected apparent magnitude of supernovae, 
and ($\lambda_A$, $\sigma_A$), ($\lambda_B$, $\sigma_B$) would correspond to models of two different SN~Ia 
populations with different absolute magnitudes and intrinsic dispersions, 
each propagated through the same cosmological model for the luminosity distance modulus.


\section{Generating Mock Data Sets}\label{sec:datasets}

We begin exploring a two-population GMM for SNeIa by generating a sample of mock SN~Ia data sets from Eq.~\ref{eq:GMM_prob}.
 We represent the difference in the two populations as a difference in absolute magnitude $M_X$ for $X = A$ or $B$ populations.  
The parameters in Eq.~\ref{eq:GMM_prob} can thus be redefined as:  $\lambda_A \rightarrow M_A$, $\lambda_B \rightarrow M_B$. 
While we will discuss absolute magnitude distributions in this section in order to emphasize the different populations, later we will consider fitting the mock data as ``observed" apparent magnitudes. 
 We define the relative mean magnitude shift between the populations such that $\dM \equiv M_A - M_B$ and re-parameterize $M_B$ in terms of $M_A$ and $\dM$ as $M_B = M_A - \dM$. 
The relative magnitude difference \dM~is thus applicable to either absolute or apparent magnitude, and the overall normalization of the absolute magnitude -- which is generally marginalized over -- is absorbed into one term for both populations.
 The variance of each population ($\sigma_{X}^2$) is defined as $\sigma_{X}^2 = \sigma_{{\rm int}, X}^2 + \sigphot^2$ including the intrinsic dispersion of the population  $\sigma_{{\rm int}, X}$ and the dispersion introduced from observational errors $\sigphot$. 

\begin{figure}
\epsscale{1.2}
	\plotone{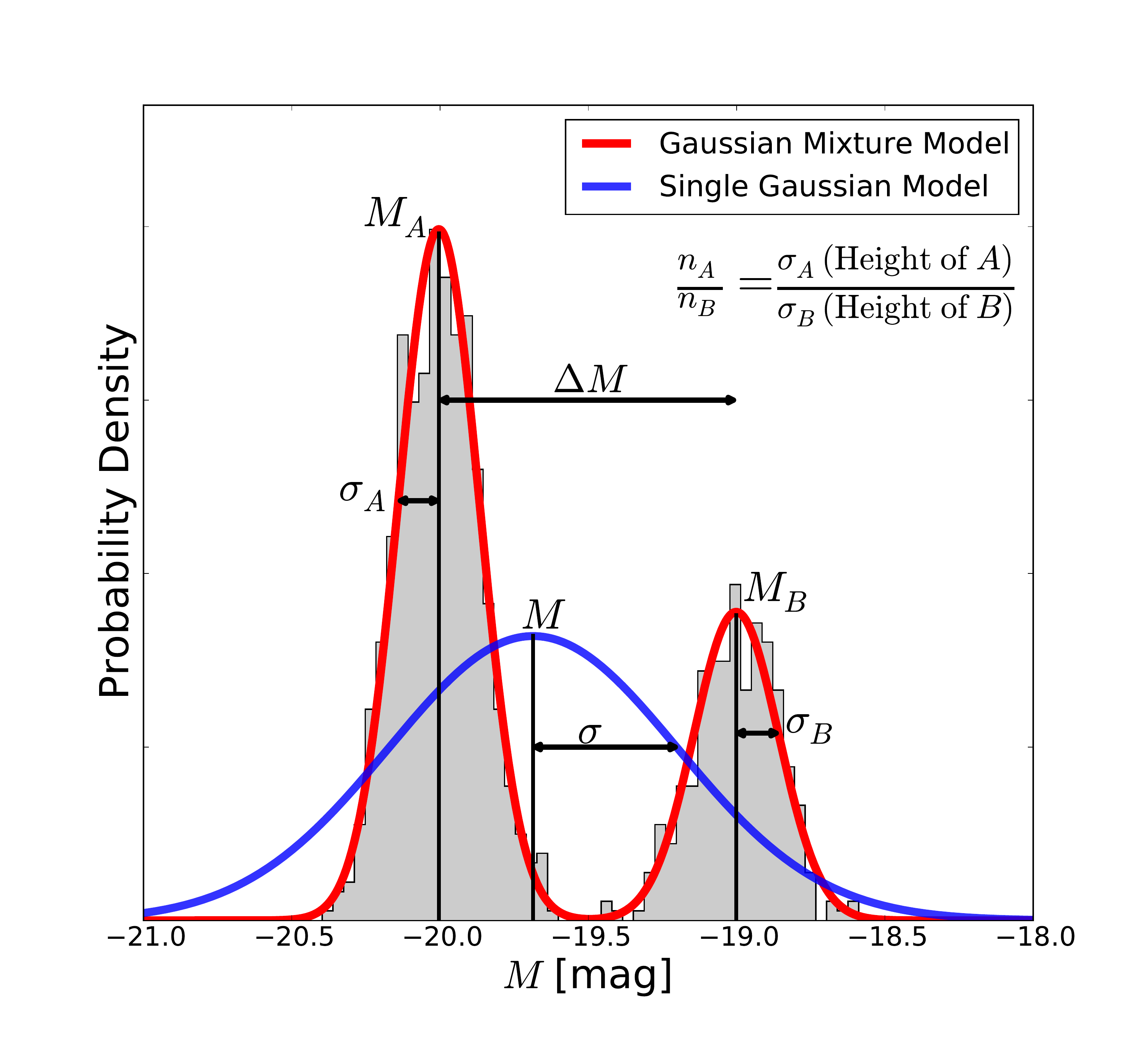}
	\caption{A histogram of mock supernovae with two populations are shown in grey and fit with a GMM and SGM. $M$ and $\sigma$ are the peak and dispersion from a SGM. The GMM model includes the location of both peaks, $M_A$ and $M_B$, and the dispersion of both populations, $\sigma_A$ and $\sigma_B$. $\dM$ is the different between $M_A$ and $M_B$ in magnitudes.  
The relative number of SNeIa in each population is $n_A/n_B$ (where $n_A+n_B=1$).  
In this example, $\dM = 1.0$~mag and $n_A = 0.7$.
}
	\label{fig:SG_GMM}
\end{figure}

Figure~\ref{fig:SG_GMM} illustrates graphically the 
five parameters of our two-population GMM: $M_A, \dM, \sigma_A, \sigma_B,$ and $n_A$ and the two parameters of a single-Gaussian model (SGM): $M$ and $\sigma$ fit to the GMM-generated data.
For visual clarity, this example has $n_A = 0.7$ and shows an extreme shift of $\dM = 1.0$~mag.  We expect realistic models to be on the order of $\dM\lesssim0.1$~mag.

 We simulate mock data sets assuming the peaks of the populations average to the estimated value of $M$ such that $(M_A + M_B) / 2 = -19.5$ mag 
with intrinsic dispersions of $\sigma_{{\rm int}, X}=0.1$~mag and $\sigphot = 0.1$~mag for both populations.
$\sigphot$ was chosen to reflect the observational error that JLA achieved ($\sim0.115$ mag).
The supernovae are constrained to a redshift range of $0.05 < z < 1.5$ to cover the low redshift anchors and the high redshift cosmology probes. 

Because host galaxy properties are on average different at $z\sim0$ and $z\sim1$, it becomes sensible to explore the possibility of redshift evolution between the relative number of SNeIa in each population.  
As a toy model we simulate a redshift dependence of the relative normalizations by having the populations evolve linearly in redshift: $n_A = n_{A, 0}' z + n_{A,0}$. 
 Where $n_{A,0}$ is  $n_A(z)$ evaluated at $z = 0$ and $n_{A,0}'$ is the first derivative of $n_A(z)$ evaluated at $z = 0$.
We then impose boundary conditions such that the total population of supernova is dominated by a single population at the lowest redshift $n_A ( z_{\rm min} = 0.05) = 1$ 
and the other population dominates the total population at the highest redshift $n_A (z_{\rm max} = 1.5) = 0$ 
to get  $n_{A,0} = 1.003$ (no units) and $n_{A,0}' = -0.627$ in units of 1/redshift. 
The two populations have an equal number of supernovae at $z=0.775$ as set by the slope and intercept of $n_A(z)$.
This value is derived only from relative normalizations and is independent of other supernova population parameters.

A linear evolution with redshift is an overly simplistic model.
 The evolution of multiple populations or other astrophysical systematics will likely be a smooth, potentially monotonic, function of redshift.  
 While a power law or logarithmic function might suggest itself as a good model for a variety of phenomena, 
 a linear dependence is at least a reasonable description of a function for which we have a strong bias that should be varying slowly. 
 As such, it is informative to explore a linear model, which is likely to capture a significant amount of the overall trend of the true astrophysical systematics.
In \cite{Greggio08}, Figure 7 (top panel) shows the relative rates of the single degenerate channel versus double degenerate channels as a function of redshift. 
These are clear parallels to our relative population parameters, and one of the models shows a linear trend.
The modeling of SNeIa progenitors is still incomplete and different models can provide drastically different rates.  
The GMM does not rely on a linear model for the evolution of the relative populations and can easily be constructed with different forms such as a power law or logarithmic function.

We randomly draw a redshift from a uniform distribution in the range $0.05 < z < 1.5$, then generate a GMM PDF corresponding to that redshift, and randomly draw an absolute magnitude from that PDF. 

\begin{figure*}
	\plotone{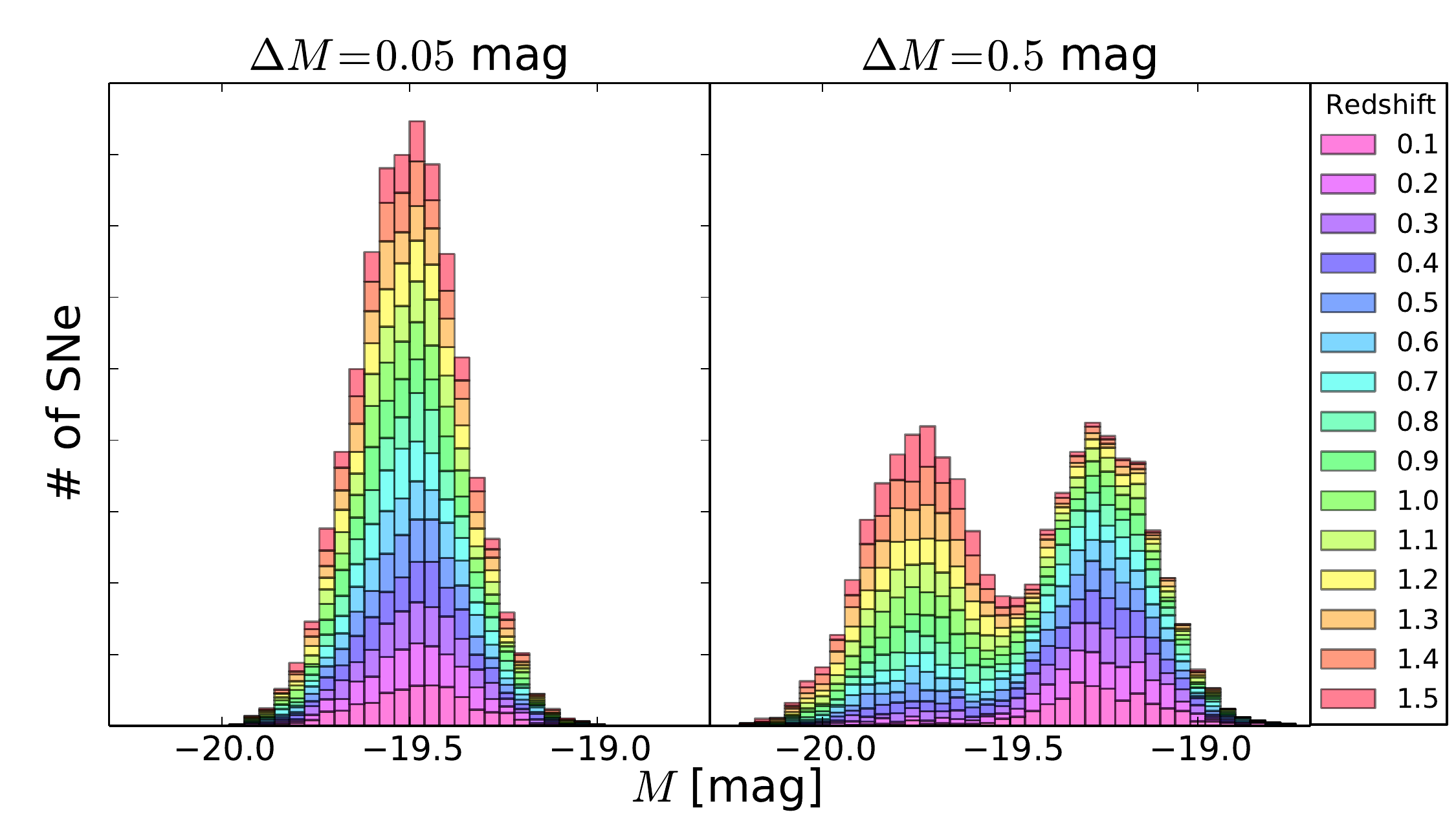}
	\caption{Absolute magnitude distribution of 10,000 mock SNeIa separated into 15 redshift bins denoted by color. A histogram is generated at each redshift then stacked upon the previous redshift's histogram. \textit{Left:} A small separation of $\dM = 0.05$~mag is a subtle shift. \textit{Right:} A exaggerated separation of $\dM = 0.5$~mag makes the evolution visually obvious.}
	\label{fig:colorhist5}
\end{figure*}

Figure~\ref{fig:colorhist5} illustrates how the absolute magnitude distribution of SNeIa evolves with redshift for two different $\dM$s. 
While the redshift evolution is a small effect for small $\dM$, the shift between different populations becomes understandably more clear when $\dM = 0.5$~mag.

We generate 108 different data sets with a range of number of supernovae in each set: $N = 100,\ 1000,\ 2500,\ 10000$ and a range of shifts between the two supernovae populations: $\dM = 0.0,\ 0.01,\ 0.02,\ 0.05,\ 0.1,\ 0.2,\ 0.3,\ 0.4,\ 0.5$~mag. 
Each permutation of $N$ with $\dM$ is performed three times to help average over random fluctuations in the data sets.
The number of supernovae correspond to a small sample, the order of current data sets ($1000$), and the expected yields from WFIRST-AFTA ($2500$) and LSST
 ($10,000$).\footnote{Current estimates of cosmologically useful SNeIa from LSST range from 10,000s to 100,000.  We have chosen a very conservative value here.}
$\dM = 0.0$~mag is consistent with a single Gaussian population while $\dM = 0.1$~mag is close to the number quoted from \cite{Rigault15} for the difference in brightness between supernovae located in active versus passive local environments.  
Though this framework is discussed with a specific systematic as an example motivation, it is general and can be applied to any systematic that can be described by an effective distribution in the likelihood. 

In order to use apparent magnitudes instead of absolute magnitude,
we add the cosmological distance modulus $\mu (z; \Omega_M, w)$ to produce an apparent magnitude ($m$).
We chose our default cosmology to be that of WMAP9 with $\Omega_M=0.2865$, $\Omega_L=0.7134$, $w=-1$, $H_0=69.32$~km~Mpc$^{-1}$~s$^{-1}$ \citep{WMAP9}. 
We do not simulate a distribution of stretch and color or the resulting correction process.  
This process is thus rather generically applicable to any luminosity distance indicator 
with no particular restriction to SNeIa beyond the parameters chosen for the GMM. 

In the present work, we also neglect the effects of gravitational lensing on SN~Ia analyses. 
Though the dispersion induced by lensing may be non-negligible in forthcoming analyses \citep{zentner_bhattacharya09}, lensing 
does not shift the average brightness (setting aside observational selection effects for the moment)
and is unlikely to bias cosmological results \citep{helbig15}. 
We defer a more complex analysis including lensing to future work.


\section{Methods}
\label{sec:methods}


\subsection{Markov Chain Monte Carlo}

We use standard Markov Chain Monte Carlo (MCMC; \citealt{Metropolis53}) techniques to fit for model parameters. In particular, we utilize the affine-invariant ensemble sampler from \cite{Goodman-Weare10} and implemented in \texttt{python} with \texttt{emcee} \citep{emcee}. We test the convergence of our chains by checking that the autocorrelation of points sampled from the posterior approaches zero for large lags \citep{Box76}.

The likelihood including cosmology used for the MCMC analysis is defined as 
\begin{eqnarray}
\mathcal{L} &= \sum\limits_{i=1}^N \ln & \left[ \frac{n_A(z)}{\sqrt{2 \pi \sigma_A^2}} \exp \left(\frac{-(m_i - m_A)^2}{2 \sigma_A^2} \right) + \nonumber  \right. \\
  & & \left. \frac{(1-n_A(z))}{\sqrt{2 \pi \sigma_B^2}} \exp \left(\frac{-(m_i - m_B)^2}{2 \sigma_B^2} \right) \right] 
\label{eq:GMM_LL}
\end{eqnarray}
where:
\begin{itemize}
	
	\item $N$ is the number of supernovae in the mock data set;
	
	\item $n_A(z)$ is the relative normalization of population~A,
	   \begin{eqnarray*} 
	   n_A (z) = n_{A,0}' z +  n_{A,0}\ \ ; 
	   \end{eqnarray*}
	  
	\item $\sigma_{X}$ is the standard deviation of the two populations such that
	    \begin{eqnarray*} 
	    \sigma_{X}^2 = \sigphot^2 + \sigma_{ {\rm int}, X}^2 \quad \text{where $X = A$ or $B$};
	    \end{eqnarray*}
	   
	\item $m_i$ is the generated ``observed" apparent magnitude for supernova $i$ in the mock data set;
	
	\item $m_A$ and $m_B$ are predicted apparent magnitudes based on cosmological parameters through the Hubble constant-free luminosity distance, 
		\begin{eqnarray*}
			 m_A  &=& 5 \log({\DL}) + \mathcal{M}_A   \quad     \\
			 & & \text{where} \quad \mathcal{M}_A  =  25 - 5 \log{H_0} + M_A \\
			 & & \quad \mathrm{and} \\
		 	m_B &=& 5 \log({\DL}) + \mathcal{M}_B  \quad   \\
		  	& &  \text{where} \quad \mathcal{M}_B = \mathcal{M}_A - \dM.
		\end{eqnarray*}
	
\end{itemize}

We assume a flat universe ($\Omega_M+\Omega_{\rm \Lambda}=1$) and fit for the matter density $\Omega_M$ and the dark energy 
equation of state parameter $w$. In the case of the GMM fits, we also fit for six nuisance parameters: $\mathcal{M_A}, ~\dM, ~\sigma_{{ \rm int},A}, ~\sigma_{ {\rm int}, B}, 
 ~n_{A,0}' $ and   $~n_{A,0}$ 
which incapsulate the information about the underlying SN~Ia populations. 
However, since we used the Hubble constant free luminosity distance, we must still specify $H_0$ 
to completely describe the underlying populations. 

In addition to our GMM analysis, we also fit each data set using a single-Gaussian model (SGM) for the underlying SN~Ia 
population; these fits have just two nuisance parameters: $\mathcal{M}$ and $\sigma_{\rm int}$. 

 For all parameters we use the flat priors defined in Table~\ref{table:prior} and an extra prior in the GMM on the combination of 
 $n_{A,0}' $ and   $n_{A,0}$ 
 such that $ 0 \leq n_A(z) \leq 1$.
 
\begin{deluxetable}{ccccccc}
\tablecaption{Flat Priors}
\tabletypesize{\footnotesize}
\tablewidth{0pt}
\tablecolumns{12}
\tablehead{ \colhead{$\Omega_M$}   & \colhead{$w$} &  \colhead{$\mathcal{M}/\mathcal{M}_A$} &\colhead{$\Delta M$}  & \colhead{$\sigma_{\text{int}, X}$}  & \colhead{$n_{A,0}'$}  & \colhead{$n_{A,0}$} }
\startdata
[0,1]  & [-3,1]  &  [-10, 5]  & [0, 5]  & [0.0, 0.3]  & [-1,0]  & [0, 2]  \\
\enddata
\label{table:prior}
\end{deluxetable}

\begin{figure*}
	\plottwo{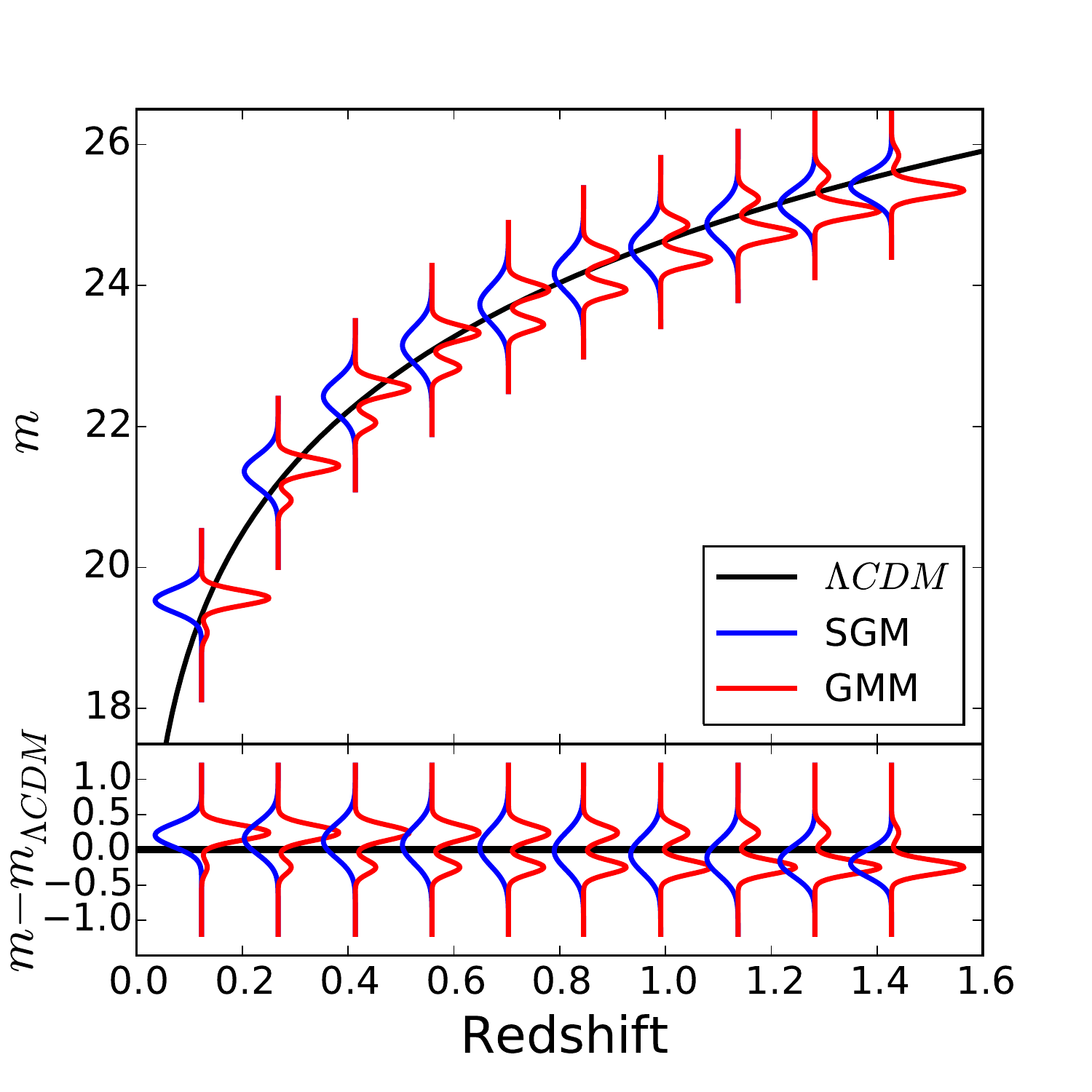}{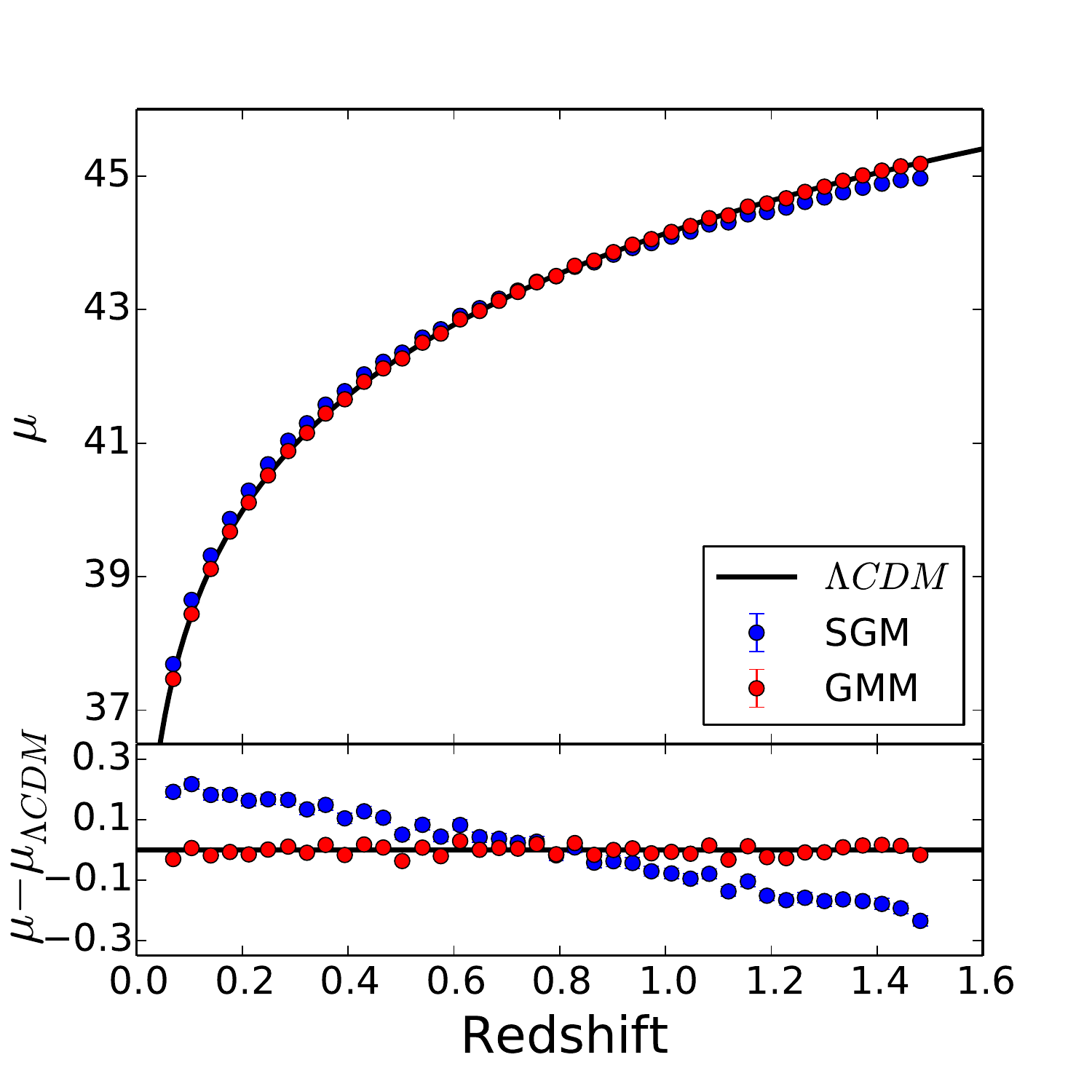}
	\caption{
	{\em Left:}
	Apparent magnitude versus redshift (top panel) and the Hubble residual (bottom panel) for parameter inferences using a GMM with $\dM = 0.5$ mag and $N=10,000$ SNeIa. 
	The black line cutting across the data is the expected magnitude redshift relation in our fiducial $\Lambda $CDM cosmology. 
	At each of ten evenly-spaced redshift bins the blue (left-directed) curves show the PDFs of $m$ inferred from a SGM fit to the GMM mock data while the red (right-directed) curves 
	show the PDFs of $m$ inferred by fitting the data with a GMM. 
	Clearly the inferred $m$ are biased in the SGM fits to the GMM mock data. 
	{\em Right:}
	Luminosity distance modulus versus redshift (top panel) and residual (bottom panel) for $\dM = 0.5$ mag with  $N=10,000$. 
	The line is the distance modulus calculated from $\Lambda$CDM.  
	The data points are the mock data sets minus the model for absolute magnitude ($\mu = m-M_{\rm model}$) using the SGM (blue) with $M_{\text{model}} \equiv M$ and the GMM (red) with $M_{\rm model} \equiv n_A M_A + n_B M_B$ with values derived from fit models holding cosmology and $H_0$ constant.
	1~$\sigma$ error bars have been plotted but are too small to see. 
}
	\label{fig:CF}
\end{figure*}

\subsection{Model Comparison} 
\label{subsection:model_comparison}

We have introduced a GMM to treat the cosmological analyses of SN~Ia data. 
The GMM is more complex than the SGM as evidenced, in part, by the fact that the GMM has four more nuisance parameters. 
The question arises whether or not the additional complexity is demanded by the data or, in our case, by the mock data used to mimic forthcoming analyses. 
We employ three statistical tests to indicate whether or not the additional complexity is required by the data: 
the Akaike Information Criterion (AIC; \citealt{Akaike74}); the Bayesian or Schwartz Information Criterion (BIC; \citealt{schwarz1978}); and the Deviance Information Criterion (DIC; \citealt{Spiegelhalter02}). 
For a review of these three methods we refer the interested reader to \cite{Liddle07} and for a more in-depth discussion of AIC and DIC see \cite{Gelman14}.

The AIC and BIC are calculated from the maximum likelihood $\mathcal{L}_{\rm max}$, the number of model parameters $k$, and the number of data points $N$ as  
\begin{equation}
\text{AIC} = -2 \ln{\mathcal{L}_{\rm max}} + 2 k + \frac{2 k (k+1)}{N-k-1}
\label{eq:AIC}
\end{equation}
and
\begin{equation}
\text{BIC} = -2 \ln{\mathcal{L}_{\rm max}} + k \ln{N}.
\label{eq:BIC}
\end{equation}
Models with lower values of these information criteria are favored. Both the AIC and BIC penalize models with a greater number of parameters (greater $k$) because $\mathcal{L}_{\rm max}$ can only increase with increased parameter freedom, while the BIC also penalizes larger data sets (greater $N$) to reduce the risk of over fitting.

The DIC is more suited for analyses with MCMC outputs because it directly uses the resulting samples from the posterior. The DIC can be computed from these samples in the MCMC chain as
\begin{equation}
\text{DIC} = 2 \overline{D({\boldsymbol \theta})} - D(\tilde{\boldsymbol \theta}), 
\label{eq:DIC}
\end{equation}
where ${\boldsymbol \theta}$ is the set of parameters directly from the samples in the chain
(in our case these are the cosmological 
parameters $w$ and $\Omega_{\rm M}$ along with the parameters of either the SGM or GMM), 
$D({\boldsymbol \theta})$ is the deviance, 
\begin{equation}
D({\boldsymbol \theta}) = -2 \ln{\mathcal{L}({\boldsymbol \theta})} + C, 
\label{eq:Deviance}
\end{equation}
$\mathcal{L}({\boldsymbol \theta})$ is the likelihood evaluated at parameters ${\boldsymbol \theta}$, and $C$ is a normalizing constant that cancels when comparing different models. 
$\overline{D({\boldsymbol \theta})}$ is the average of the deviance evaluated at each step in the 
chain and $D(\tilde{{\boldsymbol \theta}})$ is the deviance evaluated at the mean, median, or some other summary 
point in parameter space $\tilde{\boldsymbol \theta}$. 
In our samples, we find that the median is a better representation 
of our data because many of the posterior distributions are non-Gaussian, which can result in a mean value strongly influenced by tails.


\section{Results}\label{sec:results}


\subsection{An Illustration of Parameter Bias}

We illustrate the potential for bias in the inferred cosmological parameters due to multiple SN~Ia populations by first 
presenting Hubble diagrams. 
We consider data generated from an underlying GMM but fit with both a SGM likelihood 
and a GMM likelihood. 
The fit using a SGM likelihood function is intended to mimic an analysis in which there is 
no mechanism to account for two distinct populations. 

Figure~\ref{fig:CF} shows the results of a comparison between a SGM and GMM analysis 
using one data set with an exaggerated shift in the magnitude difference between the two populations, 
$\dM = 0.5$~mag. We use this large shift here for illustrative 
purposes and more realistic values are $\dM \lesssim 0.1$. 
The upper panel of the left figure in Figure~\ref{fig:CF} shows, 
within ten evenly-spaced redshift bins, the PDF of apparent magnitude inferred from both the SGM and 
GMM fits to the underlying, multi-modal, GMM mock data.
The parameters of these PDFs are determined by the fits through the MCMC 
process described in Section~\ref{sec:methods} with the cosmological parameters held constant for simplicity. 
The SGM was fit at each redshift bin while the GMM was fit using all the data at once to constrain the parameters of 
redshift evolution. 
The peak of the SGM PDF in the residual ($m_{\rm data} - m_{\Lambda{\rm CDM}}$)
exhibits a linear evolution getting brighter as redshift increases,
which is the result of the redshift evolution in the data set.

The right plot in Figure~\ref{fig:CF} shows the same data set and MCMC fit (with cosmology constant) converted into distance modulus versus redshift. 
Simply subtracting the absolute magnitude derived from the MCMC fit of the mock data yields this information. 
The absolute magnitude for the SGM can be taken straight from the chains ($M_{\rm SGM} = M$); however, 
the absolute magnitude for the GMM is a function of redshift and multiple fitted parameters ($M_{\rm GMM}(z)  = n_A(z) M_A + n_B(z) M_B$).
The inferred SN~Ia population parameters $M$ and $\sigma$ for the SGM have no way to account for the relative shift between the two SN~Ia populations as a function of redshift and so the SGM fits show a systematic, redshift-dependent deviation in the distance modulus as a function of redshift. 
Notice that the mock GMM data set was generated such that at $z=0.775$, the two populations have an equal number of SNeIa and, as expected, $M_{\rm SGM} = M_{\rm GMM}$ at $z=0.775$. 
The population parameters are recovered well for the GMM fit and there is clearly no bias in this case. 


\subsection{Cosmological Parameters}

From the perspective of exploiting SNeIa as a probe of cosmology, the greatest concern caused by 
multiple populations of SNeIa is that insufficiently accurate modeling of the multiple populations will 
lead to biased cosmological parameters. Exploring this possibility is the primary purpose of this 
paper. 
To explore the potential importance of multiple 
SN~Ia populations on cosmology, we fit each of the 108 mock data sets described in Section~\ref{sec:datasets} 
for the cosmological parameters, $\Omega_{\rm M}$ and $w$, and SN~Ia population parameters simultaneously. 

\begin{figure}
	\plotone{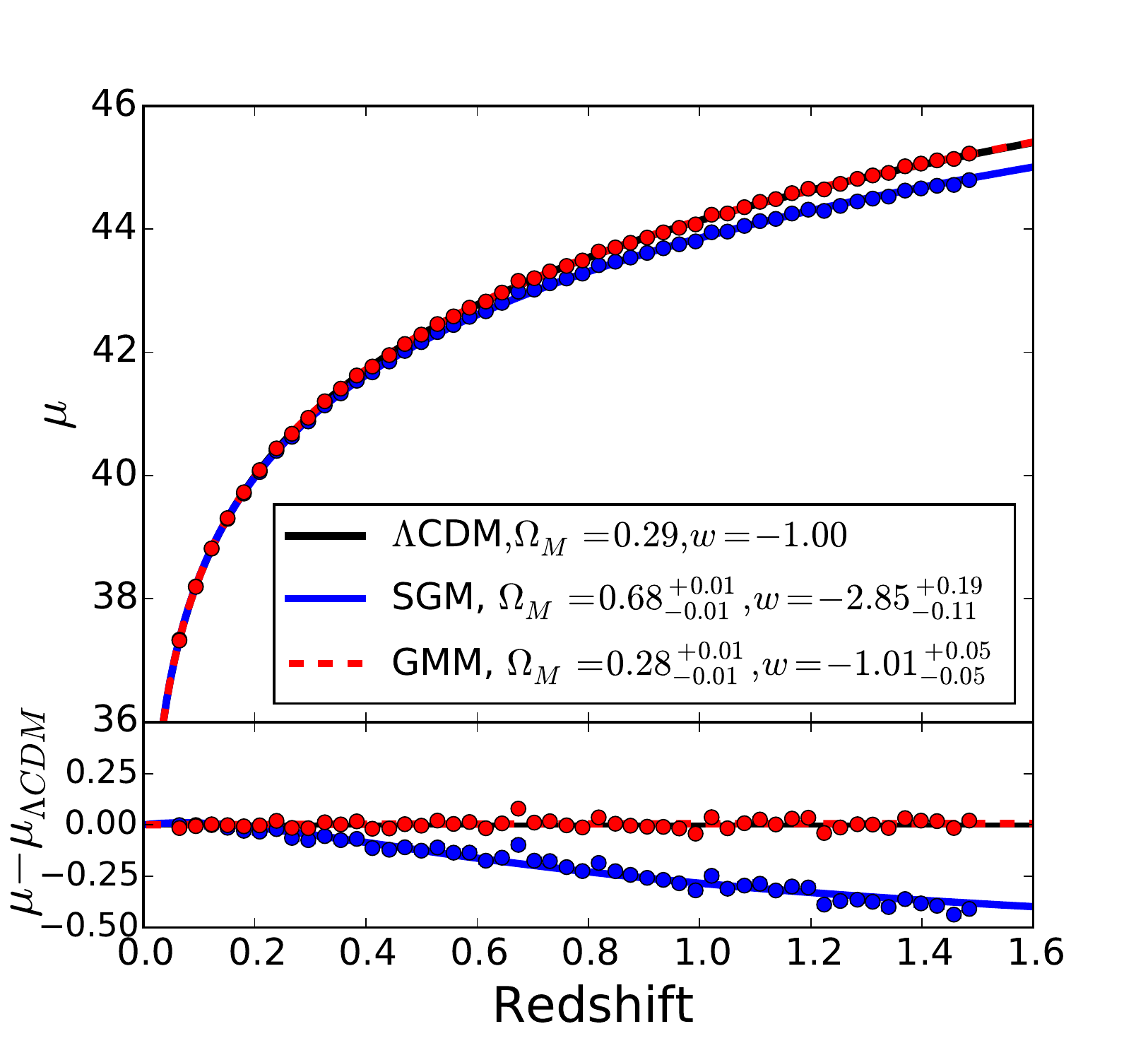}
	\caption{
	Hubble constant-free luminosity distance modulus versus redshift (top panel) and residual (bottom panel) for $\dM = 0.5$ mag with  $N=10,000$. 
	The lines are the distance modulus calculated from cosmology derived from the MCMC fits.  
	The data points are the mock data sets minus the model for absolute magnitude ($\mu = m-M_{\rm model}$) using the SGM (blue) with $M_{\text{model}} \equiv M$ and the GMM (red) with $M_{\rm model} \equiv n_A M_A + n_B M_B$ with values derived from fit models.
	The data points correspond in color to their model. 
	1~$\sigma$ error bars have been plotted but are too small to see. 
}
	\label{fig:HD}
\end{figure}

Figure~\ref{fig:HD} displays the Hubble diagram inferred from both SGM and GMM fits to a GMM model from a single data set with 
$N=10,000$ SNeIa and an extreme value of $\dM=0.5$~mag. This large value of $\dM$ is used to produce this 
figure only because it has the pedagogical value of making the influence of the two-populations model on inferred 
cosmology obvious. Clearly the GMM fits yield an unbiased Hubble diagram and we infer unbiased values of 
both $\Omega_{\rm M}$ and $w$. 

\
 
\begin{figure*}
	\plotone{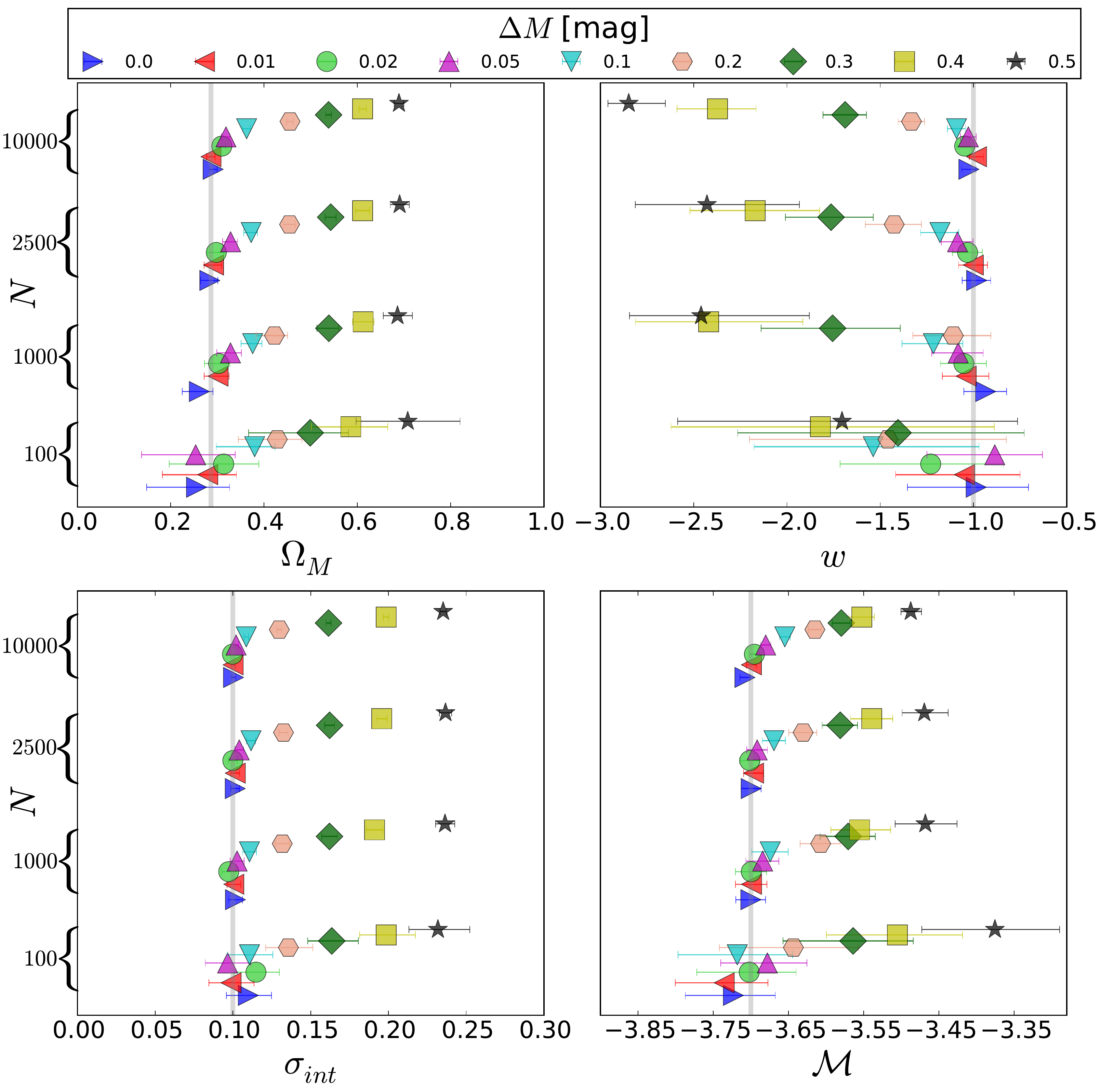}
	\caption{The median and $68\%$ confidence region from the MCMC analysis using a SGM likelihood plotted within the limits of each prior (except $\scriptM$). 
	Each $\dM$ is given its own color and shape. 
	To avoid overlap in the error bars, we present the increasing $\Delta M$ values with a small offset in ordinate value within each $N$ bracket. 
	The grey vertical line in each plot marks the fiducial value of that parameter.}
	\label{fig:SG_estp}
\end{figure*}

On the other hand, the SGM fits to the GMM produces a biased inferred Hubble diagram and biased inferences 
for the cosmological parameters. Compare Fig.~\ref{fig:HD} to the right plot of Fig.~\ref{fig:CF}. Notice that the results of 
the two fits no longer cross near $z=0.775$ once cosmological parameters are fit simultaneously with SN~Ia 
population parameters. The SGM fits to the GMM mock data result in cosmological parameters and SN~Ia population 
parameters that are simultaneously significantly biased. As a result, the inferred Hubble diagram differs from the 
true underlying dependence of distance modulus on redshift. Most importantly, the bias in the cosmological parameters 
is significant. We infer $\Omega_{\rm M}=0.69 \pm 0.01$ and $w=-2.85^{+0.19}_{-0.11}$ and rule out the true 
underlying cosmology with high confidence. Of course, this model with $\dM=0.5$~mag is extreme, but we will now 
move on to a discussion of inferred cosmological parameters in each of our 108 mock data sets and show that 
viable two-population SN~Ia models yield biases in cosmology that are non-negligible compared to statistical errors.

\begin{figure}
	\plotone{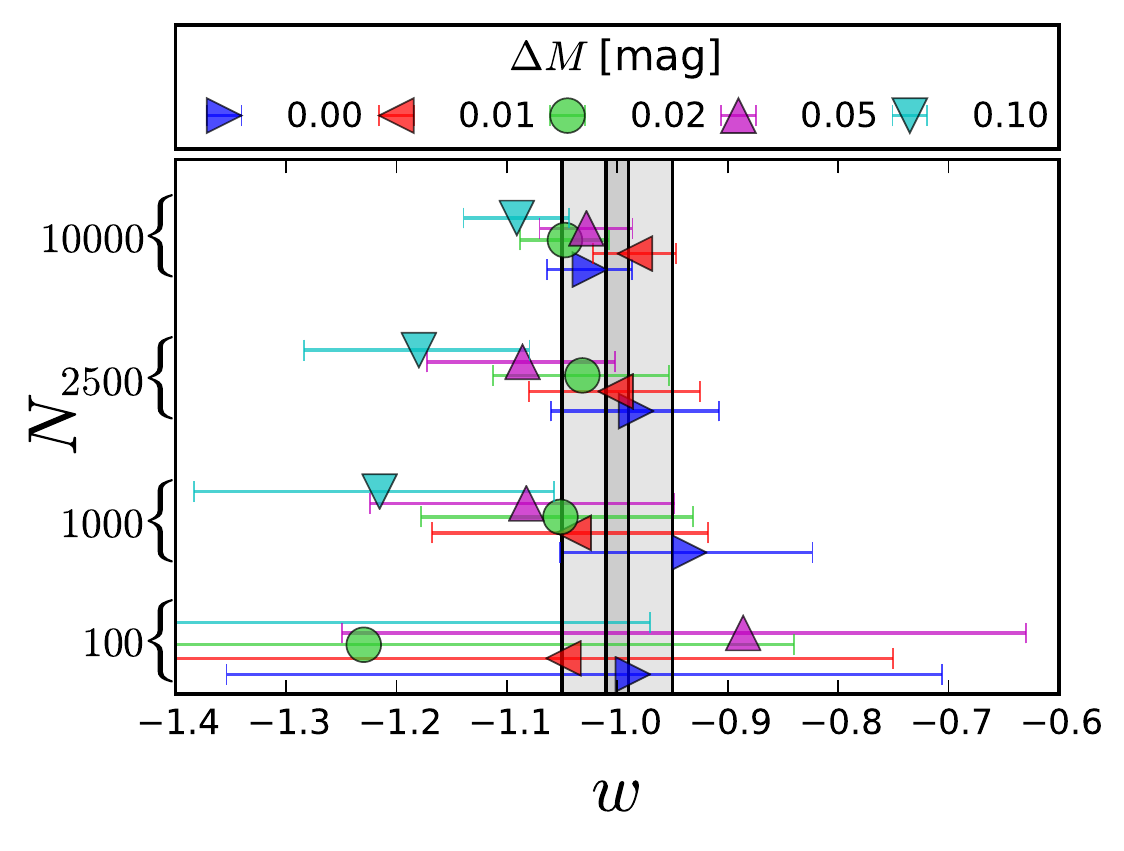}
	\caption{A close up view of the Top Right panel from Figure~\ref{fig:SG_estp} to show the induced bias on $w$. 
We here only focus on the empirically plausible values $0.0 < \dM < 0.1$~mag. 
The vertical lines indicate ranges of $\pm0.05$ and $\pm0.01$ in $w$ for reference.}
	\label{fig:w_bias}
\end{figure}

We present medians and 68\% confidence regions of the fitted parameters by combining the MCMC results from the 3 different data sets at each value of $N$ and at each value of $\dM$. 
We define the 68\% confidence region as the area contained within the 16$^{\rm th}$ and 84$^{\rm th}$ percentiles, 
which enforces an equal probability in the tails at either end of the posterior distribution. 
In order to combine the three data sets,
 we calculate the average of the medians, and 
we calculate the 16$^{\rm th}$ and 84$^{\rm th}$ percentiles as 
\begin{equation}
\label{eq:Errors}
\sigma_{\%}= \sqrt{\frac{\sigma_{\%,1}^2 + \sigma_{\%, 2} ^2  + \sigma_{\%,3}^2}{3}},
\end{equation}
where $\sigma_{\%}$ is the 16$^{\rm th}$ or 84$^{\rm th}$ percentile and $\sigma_{\%,i}$ corresponds to the 16$^{\rm th}$ or 84$^{\rm th}$ percentile calculated from the $i^{\rm th}$ data set.

Fig.~\ref{fig:SG_estp} shows the medians and 68\% confidence regions in the inferred parameters in our fits using a SGM to 
describe GMM mock data. 
As Fig.~\ref{fig:SG_estp} clearly shows, for $\dM=0$, the inferred parameters are unbiased: 
the true, underlying value of each of the cosmological parameters is inferred to within statistical precision. 
This is unsurprising. 
We have assumed that both sub-populations have the same intrinsic dispersion, so a 
model in which $\dM=0$ is tantamount to a SGM for SNeIa. 
This is nothing more than a validation of this 
procedure for a single population of SNeIa. Models with $\dM \ne 0$ correspond to GMM models. 
Both cosmological and SN~Ia population model parameters exhibit increasing biases as $\dM$ increases. Moreover, many of these biases are quite 
statistically significant suggesting that it is possible to rule out the correct underlying models due to these systematic 
errors. We note that in some cases ($\dM \gtrsim 0.4$) the inferred values of $w$ are strongly influenced by the hard prior $w>-3$ 
that we have enforced. 
Table~\ref{table:cosmo_parms} summarizes only the results for cosmological parameters with 68\% confidence regions for SGM and GMM results.

\begin{turnpage}
\begin{deluxetable*}{rrrrrrrrrrrr}
\tablecaption{The median and $68\%$ confidence region of $\Omega_M$ and $w$ for all $N$ and $\dM$.  }
\tabletypesize{\footnotesize}
\tablewidth{0pt}
\tablecolumns{12}
\tablehead{\colhead{N} & \colhead{Model}  & \colhead{$\Omega$}  & \colhead{$\dM =0.0$}  & \colhead{0.01}  & \colhead{0.02}  & \colhead{0.05}  & \colhead{0.1}  & \colhead{0.2}  &  \colhead{0.3}  & \colhead{0.4}   & \colhead{0.5} }
\startdata
100 & GMM & $\Omega_M$ & $0.234 _{- 0.099 }^{+ 0.071 } $ & $0.265 _{- 0.093 }^{+ 0.061 } $ & $0.295 _{- 0.104 }^{+ 0.075 } $ & $0.23 _{- 0.106 }^{+ 0.083 } $ & $0.356 _{- 0.087 }^{+ 0.049 } $ & $0.408 _{-0.088 }^{+ 0.056 } $ & $0.406 _{- 0.138 }^{+ 0.122 } $ & $0.428 _{- 0.126 }^{+ 0.112 } $ & $0.303 _{- 0.089 }^{+ 0.089}$ \\
\vspace*{3pt}

100 & SGM & $\Omega_M$ & $0.255 _{- 0.107 }^{+ 0.071 } $ & $0.279 _{- 0.097 }^{+ 0.062 } $ & $0.314 _{- 0.117 }^{+ 0.075 } $ & $0.253 _{- 0.116 }^{+ 0.085 } $ & $0.38 _{- 0.082 }^{+ 0.044 } $ & $0.428 _{- 0.084 }^{+ 0.053 } $ & $0.499 _{- 0.132 }^{+ 0.082 } $ & $0.586 _{- 0.084 }^{+ 0.079 } $ & $0.708 _{- 0.111 }^{+ 0.112}$ \\
\vspace*{3pt}

100 & GMM & $w$ & $-0.988 _{- 0.346 }^{+ 0.265 } $ & $-1.047 _{- 0.357 }^{+ 0.286 } $ & $-1.189 _{- 0.470 }^{+ 0.352 } $ & $-0.875 _{- 0.331 }^{+ 0.231 } $ & $-1.46 _{- 0.627 }^{+ 0.513 } $ & $-1.471 _{- 0.721 }^{+ 0.615 } $ & $-1.265 _{- 0.758 }^{+ 0.513 } $ & $-1.447 _{- 0.740 }^{+ 0.626 } $ & $-1.341 _{- 0.554 }^{+ 0.466}$ \\
\vspace*{3pt}

100 & SGM & $w$ & $-0.986 _{- 0.368 }^{+ 0.28 } $ & $-1.049 _{- 0.369 }^{+ 0.298 } $ & $-1.23 _{- 0.485 }^{+ 0.389 } $ & $-0.886 _{- 0.363 }^{+ 0.256 } $ & $-1.538 _{- 0.638 }^{+ 0.567 } $ & $-1.459 _{- 0.742 }^{+ 0.634 } $ & $-1.404 _{- 0.859 }^{+ 0.676 } $ & $-1.821 _{- 0.800 }^{+ 0.933 } $ & $-1.705 _{- 0.881 }^{+ 0.941}$ \\
\vspace*{3pt}

1000 & GMM & $\Omega_M$ & $0.247 _{- 0.040 }^{+ 0.032 } $ & $0.282 _{- 0.038 }^{+ 0.030 } $ & $0.285 _{- 0.036 }^{+ 0.028 } $ & $0.314 _{- 0.035 }^{+ 0.027 } $ & $0.363 _{- 0.035 }^{+ 0.025 } $ & $0.361 _{- 0.061 }^{+ 0.052 } $ & $0.289 _{- 0.048 }^{+ 0.040 } $ & $0.287 _{- 0.043 }^{+ 0.036 } $ & $0.295 _{- 0.041 }^{+ 0.033}$ \\
\vspace*{3pt}

1000 & SGM & $\Omega_M$ & $0.261 _{- 0.037 }^{+ 0.029 } $ & $0.301 _{- 0.030 }^{+ 0.024 } $ & $0.303 _{- 0.030 }^{+ 0.024 } $ & $0.328 _{- 0.029 }^{+ 0.023 } $ & $0.376 _{- 0.025 }^{+ 0.020 } $ & $0.422 _{- 0.039 }^{+ 0.028 } $ & $0.539 _{- 0.026 }^{+ 0.021 } $ & $0.612 _{- 0.023 }^{+ 0.023 } $ & $0.686 _{- 0.030 }^{+ 0.032}$ \\
\vspace*{3pt}

1000 & GMM & $w$ & $-0.92 _{- 0.113 }^{+ 0.108 } $ & $-1.014 _{- 0.125 }^{+ 0.117 } $ & $-1.025 _{- 0.122 }^{+ 0.116 } $ & $-1.06 _{- 0.139 }^{+ 0.129 } $ & $-1.184 _{- 0.167 }^{+ 0.155 } $ & $-0.994 _{- 0.197 }^{+ 0.173 } $ & $-0.993 _{- 0.170 }^{+ 0.152 } $ & $-1.003 _{- 0.162 }^{+ 0.146 } $ & $-1.014 _{- 0.150 }^{+ 0.140}$ \\
\vspace*{3pt}

1000 & SGM & $w$ & $-0.935 _{- 0.117 }^{+ 0.112 } $ & $-1.039 _{- 0.128 }^{+ 0.122 } $ & $-1.051 _{- 0.126 }^{+ 0.120 } $ & $-1.082 _{- 0.142 }^{+ 0.134 } $ & $-1.215 _{- 0.168 }^{+ 0.158 } $ & $-1.108 _{- 0.217 }^{+ 0.200 } $ & $-1.755 _{- 0.383 }^{+ 0.362 } $ & $-2.42 _{- 0.392 }^{+ 0.505 } $ & $-2.46 _{- 0.384 }^{+ 0.580}$ \\
\vspace*{3pt}

2500 & GMM & $\Omega_M$ & $0.267 _{- 0.028 }^{+ 0.024 } $ & $0.281 _{- 0.024 }^{+ 0.020 } $ & $0.285 _{- 0.025 }^{+ 0.020 } $ & $0.311 _{- 0.027 }^{+ 0.021 } $ & $0.36 _{- 0.023 }^{+ 0.017 } $ & $0.318 _{- 0.029 }^{+ 0.032 } $ & $0.296 _{- 0.027 }^{+ 0.024 } $ & $0.298 _{- 0.028 }^{+ 0.024 } $ & $0.239 _{- 0.032 }^{+ 0.029}$ \\
\vspace*{3pt}

2500 & SGM & $\Omega_M$ & $0.284 _{- 0.020 }^{+ 0.017 } $ & $0.292 _{- 0.020 }^{+ 0.017 } $ & $0.298 _{- 0.019 }^{+ 0.016 } $ & $0.328 _{- 0.017 }^{+ 0.015 } $ & $0.372 _{- 0.016 }^{+ 0.013 } $ & $0.455 _{- 0.013 }^{+ 0.012 } $ & $0.543 _{- 0.012 }^{+ 0.012 } $ & $0.611 _{- 0.015 }^{+ 0.015 } $ & $0.691 _{- 0.020 }^{+ 0.020}$ \\
\vspace*{3pt}

2500 & GMM & $w$ & $-0.967 _{- 0.077 }^{+ 0.073 } $ & $-0.988 _{- 0.078 }^{+ 0.075 } $ & $-1.012 _{- 0.080 }^{+ 0.077 } $ & $-1.057 _{- 0.087 }^{+ 0.083 } $ & $-1.16 _{- 0.104 }^{+ 0.100 } $ & $-1.095 _{- 0.116 }^{+ 0.100 } $ & $-1.02 _{- 0.102 }^{+ 0.094 } $ & $-0.994 _{- 0.103 }^{+ 0.098 } $ & $-0.865 _{- 0.091 }^{+ 0.083}$ \\
\vspace*{3pt}

2500 & SGM & $w$ & $-0.983 _{- 0.077 }^{+ 0.075 } $ & $-1.001 _{- 0.079 }^{+ 0.076 } $ & $-1.032 _{- 0.081 }^{+ 0.078 } $ & $-1.086 _{- 0.087 }^{+ 0.084 } $ & $-1.18 _{- 0.104 }^{+ 0.100 } $ & $-1.425 _{- 0.154 }^{+ 0.146 } $ & $-1.763 _{- 0.245 }^{+ 0.226 } $ & $-2.17 _{- 0.349 }^{+ 0.345 } $ & $-2.429 _{- 0.385 }^{+ 0.496}$ \\
\vspace*{3pt}

10000 & GMM & $\Omega_M$ & $0.285 _{- 0.013 }^{+ 0.010 } $ & $0.28 _{- 0.014 }^{+ 0.011 } $ & $0.303 _{- 0.013 }^{+ 0.010 } $ & $0.312 _{- 0.015 }^{+ 0.011 } $ & $0.321 _{- 0.031 }^{+ 0.033 } $ & $0.307 _{- 0.014 }^{+ 0.015 } $ & $0.28 _{- 0.014 }^{+ 0.013 } $ & $0.286 _{- 0.013 }^{+ 0.012 } $ & $0.283 _{- 0.012 }^{+ 0.012}$ \\
\vspace*{3pt}

10000 & SGM & $\Omega_M$ & $0.291 _{- 0.009 }^{+ 0.008 } $ & $0.286 _{- 0.009 }^{+ 0.009 } $ & $0.309 _{- 0.009 }^{+ 0.008 } $ & $0.318 _{- 0.009 }^{+ 0.009 } $ & $0.363 _{- 0.009 }^{+ 0.008 } $ & $0.456 _{- 0.007 }^{+ 0.007 } $ & $0.538 _{- 0.006 }^{+ 0.006 } $ & $0.611 _{- 0.008 }^{+ 0.008 } $ & $0.689 _{- 0.010 }^{+ 0.010}$ \\
\vspace*{3pt}

10000 & GMM & $w$ & $-1.016 _{- 0.04 0}^{+ 0.039 } $ & $-0.976 _{- 0.039 }^{+ 0.038 } $ & $-1.037 _{- 0.042 }^{+ 0.041 } $ & $-1.017 _{- 0.044 }^{+ 0.044 } $ & $-1.028 _{- 0.061 }^{+ 0.055 } $ & $-1.028 _{- 0.050 }^{+ 0.047 } $ & $-0.98 _{- 0.048 }^{+ 0.046 } $ & $-1.014 _{- 0.050 }^{+ 0.048 } $ & $-0.991 _{- 0.048 }^{+ 0.045}$ \\
\vspace*{3pt}

10000 & SGM & $w$ & $-1.025 _{- 0.039 }^{+ 0.038 } $ & $-0.984 _{- 0.038 }^{+ 0.038 } $ & $-1.047 _{- 0.041 }^{+ 0.040 } $ & $-1.028 _{- 0.042 }^{+ 0.042 } $ & $-1.091_{- 0.048 }^{+ 0.047 } $ & $-1.333 _{- 0.072 }^{+ 0.070 } $ & $-1.688 _{- 0.118 }^{+ 0.114 } $ & $-2.372 _{- 0.217 }^{+ 0.207 } $ & $-2.848 _{- 0.111 }^{+ 0.196}$ \\

\enddata
\label{table:cosmo_parms}
\end{deluxetable*}
\end{turnpage}

Fig.~\ref{fig:w_bias} is an analogous plot focusing on the inferred values of $w$, which is the primary science goal of 
dark energy probes, and observationally-plausible values of $\dM \leq 0.1$. Even in this restricted range of $\dM$ it is 
apparent that neglecting the possibility of multiple populations can lead to biases in the inferred value of $w$ that are 
non-negligible compared to the statistical errors in these parameters. This is clearly a 
challenge to precision measurements of the dark energy equation of state that must be overcome in order 
to fully exploit SNeIa. 

\begin{figure*}
	\plotone{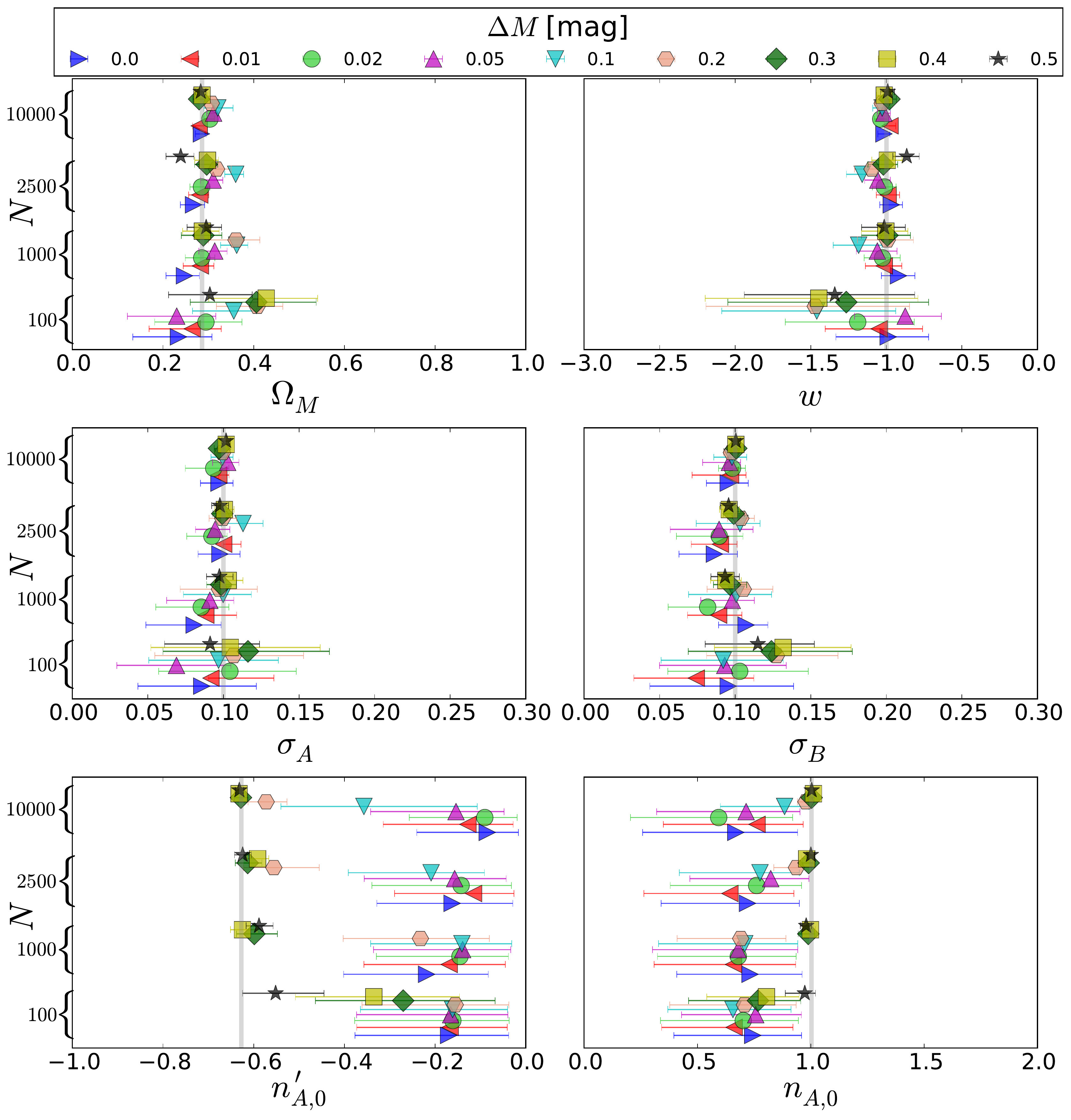}
	\caption{
The GMM fit results for the same simulations as in Figure~\ref{fig:SG_estp} and displayed with the same ordinate offsets.
The GMM model correctly recovers the fiducial cosmology and accounts for the multiple input populations.
The multiple-population parameters ($\sigma_A$, $\sigma_B$) are not well constrained for small $\dM$, 
and the normalization factors ($n_{A,0}$ and $n_{A,0}'$) are even clearly biased at low $\dM$
due to the reduced leverage they have on the output.  
But the resulting cosmological parameters are well-constrained when marginalizing over the multiple-population parameters.
}
	\label{fig:GM_estp}
\end{figure*} 

 \begin{figure*}
	\plotone{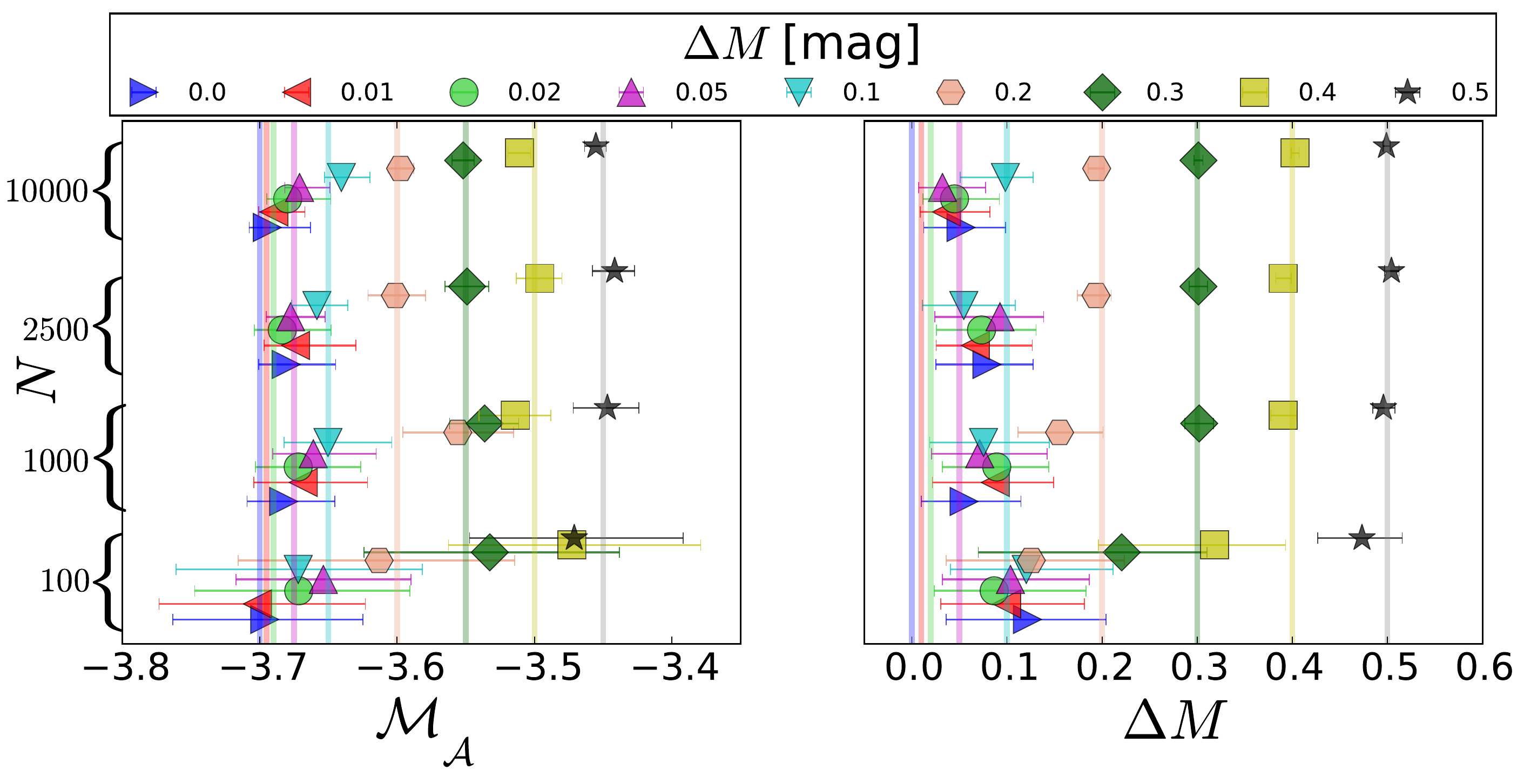}
	\caption{
	The GMM fit results for parameters dependent on  $\dM$ from the same simulations as in Figures~\ref{fig:SG_estp} and~\ref{fig:GM_estp} and displayed with the same ordinate offsets.
	Each specific $\dM$ has a vertical line denoting the different fiducial values. 
	While Figure~\ref{fig:SG_estp} and Figure~\ref{fig:GM_estp} are plotted over the entire range of the uniform prior, in this figure we focus on a range much smaller than the prior to show the detailed effect.
	}
	\label{fig:GM_estpDM}
\end{figure*}


In comparison, the inferred parameters in the GMM model fits to the GMM mock data can be seen in 
Figure~\ref{fig:GM_estp} and Figure~\ref{fig:GM_estpDM}. In {\em all} such cases we recover the 
correct cosmological parameters to within statistical precision. Indeed this is not entirely surprising 
because this is now a fit with a model that correctly describes the mock data. Indeed, we are able to 
infer all of the model parameters in an unbiased way except for  
$n_{A,0}'$ and $n_{A,0} $ 
when $\dM \lesssim 0.2$~mag. 
The fiducial values are recovered within 
the 99$\%$ confidence region for the intercept 
$n_{A,0} $ 
and within $\sim 99.9\%$ confidence region for the slope 
$n_{A,0}'$. 
It is clear that $n_{A,0}'$ and $n_{A,0} $ 
are biased in Figure~\ref{fig:GM_estp} in a way that favors 
less redshift dependence (smaller $n_{A,0}'$) except for large shifts in peaks of the two populations. 
Even though these parameters are biased, they do not introduce an increase in the variance of cosmological parameters. 
This counter intuitive result can be explained through Figure \ref{fig:triangle}, which shows that the posterior distributions of the population versus cosmological parameters are parallel to the population parameters meaning they have little to no degeneracy with 
cosmological parameters. 
When $\dM$ is sufficiently small, data with the precision and size of our mock data 
sets cannot clearly distinguish the two peaks because the separation between the 
peaks is comparable to the dispersion in any one of the sub-populations. 
It is important to note that cosmological parameters 
can be strongly biased despite the fact that a fit to the underlying data {\em cannot} 
clearly distinguish the two populations. 
This is relevant to the results of the following 
subsection.

Clearly, an underlying model in which $\dM=0$ and $\sigma_A = \sigma_B$ can be described by a SGM with no bias. 
Using a GMM model to describe such data introduces additional parameters and necessarily leads 
to less restrictive constraints on the cosmological parameters of interest. This loss in precision is the cost of 
using a model with the parameter freedom to account for the possibility of multiple SNeIa sub-populations. 
For a data set with the precision expected of $N=2500$ ($N=10,000$) SNeIa, the loss of precision in 
$\Omega_{\rm M}$ is $\sim 20\%$ ($\sim 25\%$) while the loss of precision in $w$ is approximately 
$\lesssim 1\%$ ($\lesssim 3\%$). This very moderate cost in precision greatly outweighs the 
potential $\sim 2\sigma$ statistical error that can be induced by treating a two populations of 
SNeIa with $\dM \sim 0.1$ as a single populations (see Table~2). 
This finding reaffirms that the precision does not significantly decrease when these population parameters are added to the model.


\begin{figure}
	\plotone{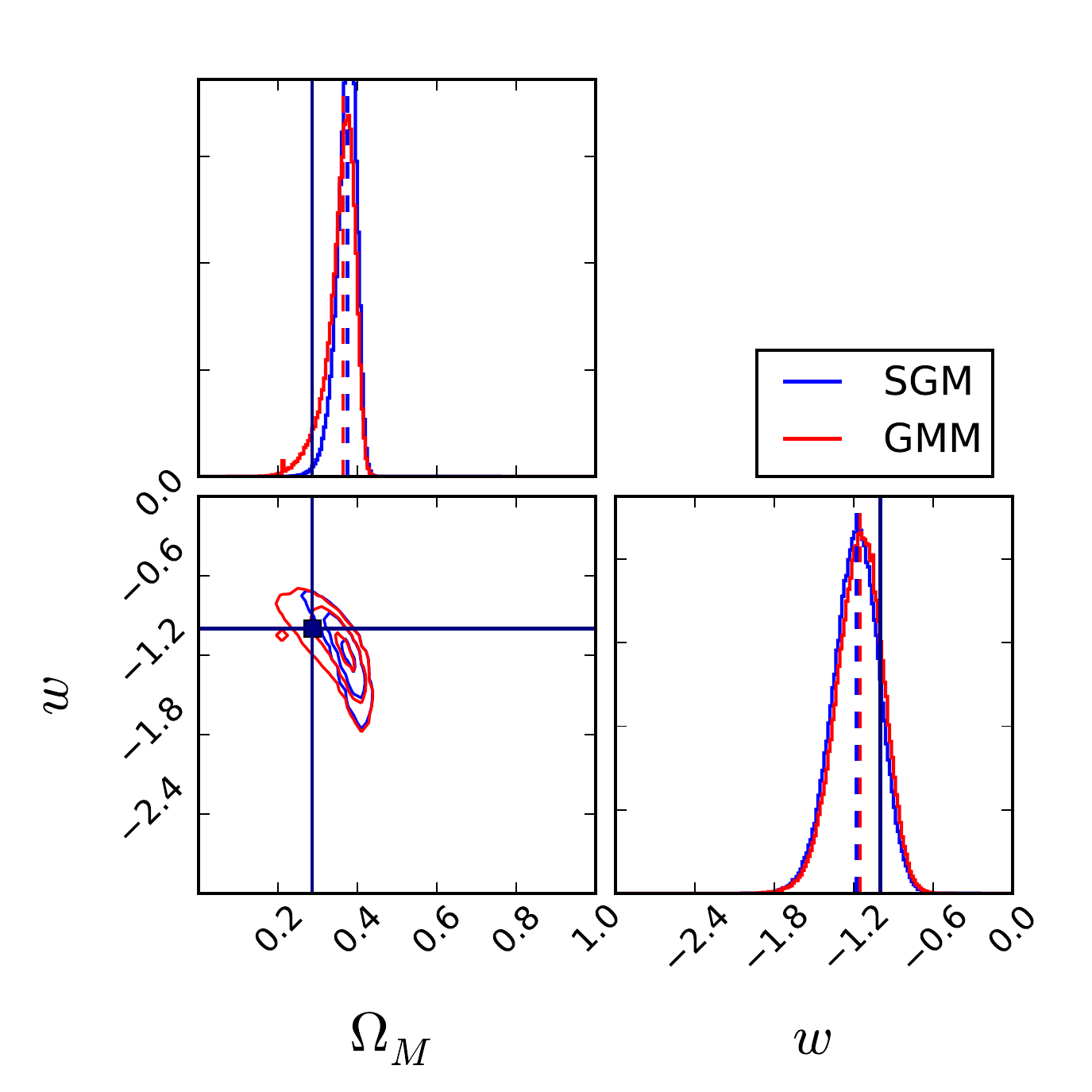}
	\caption{Cosmological contours for $\dM = 0.1$~mag and $N=1000$. Made with \texttt{triangle.py} from \cite{trianglepy}. 
	The blue contours and histograms correspond to the mock data being fit with SGM likelihood and the red contours and histograms correspond to the GMM likelihood. 
	The dashed lines are the medians of the populations. 
	The dark navy lines are the fiducial values. 
	The GMM is less precise but also less biased.} 
	\label{fig:contour} 
\end{figure} 


Figure \ref{fig:contour} shows the cosmological parameter posteriors from one data set for the interesting case of 
$N=1,000$ and $\dM = 0.1$~mag. 
These numbers are interesting since the JLA has  $\sim 1000$ SNeIa, 
and the current estimated discrepancy in Hubble residuals is equivalent to $\dM \sim 0.1$~mag. 
The contours continue to show that the GMM is less biased but also slightly less precise. 
These are not large offsets, but it could lead to a small systematic error in the next stages of observational cosmology. \\


\begin{figure*} 
	\plotone{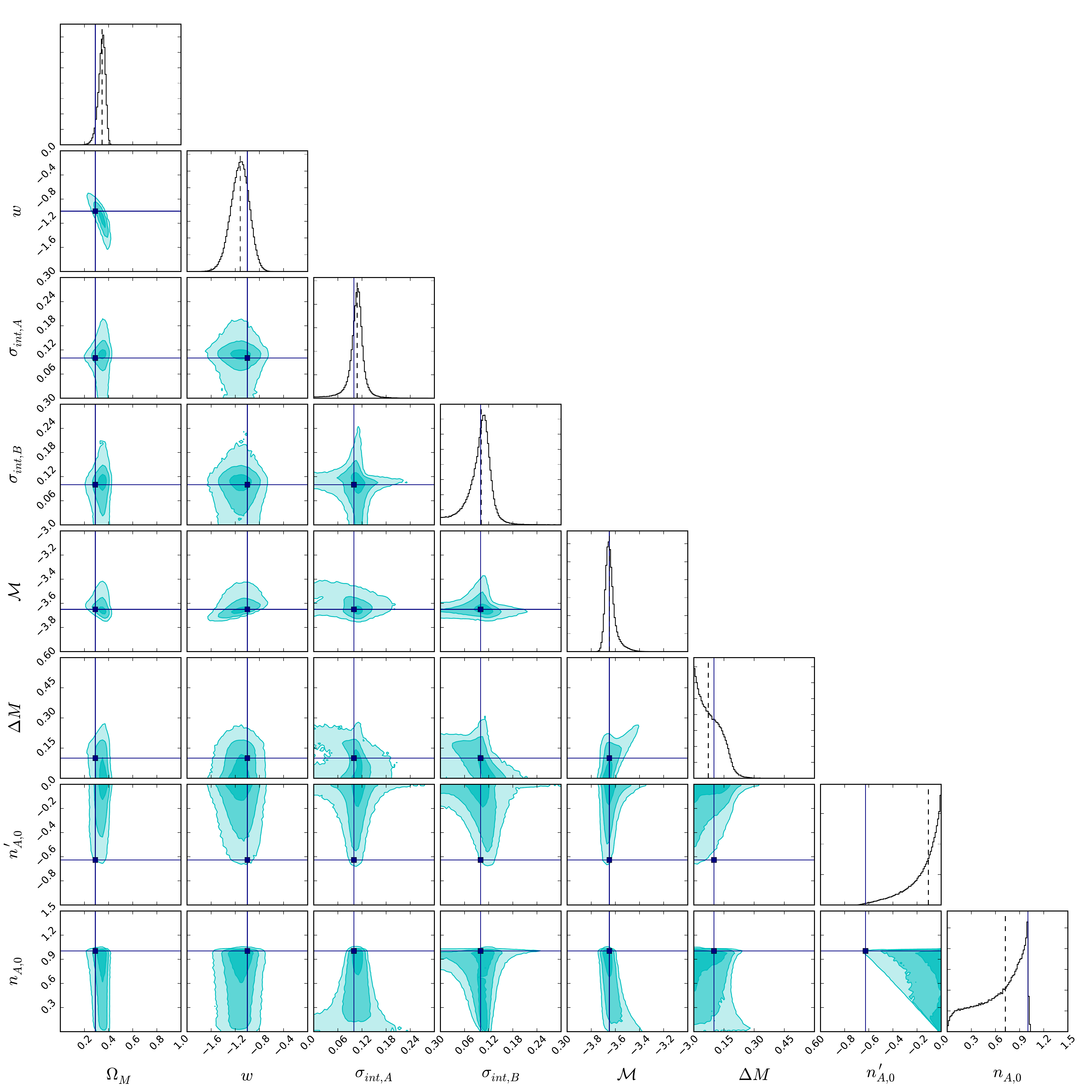}
	\caption{Full triangle plot of posterior distributions from a GMM likelihood for a single data set with $N=1,000$ and $\dM = 0.1$~mag. Made with \texttt{triangle.py} from \cite{trianglepy}}
	\label{fig:triangle} 
\end{figure*} 


\subsection{Model Selection}

To determine if the additional complexity of a given model is demanded by the (mock) data, 
we use the information criteria described in Section~\ref{subsection:model_comparison}. 
In order to compare two models, one can compute the information criteria for each and take 
the difference between the two results. 
For example, if we compute the AIC for each model, 
we would compute $\Delta \mathrm{AIC} = \mathrm{AIC}_{\rm GMM}-\mathrm{AIC}_{SGM}$ 
where $\mathrm{AIC}_{\rm GMM}$ is the value of AIC in the GMM model and likewise for $\mathrm{AIC}_{\rm SGM}$. 
We follow this convention, subtracting the SGM criteria from the GMM criteria, so that 
lower values of the difference between information criteria (IC) favor the GMM model. 
With these conventions, any change in information criteria (generically, $\Delta$ IC) will 
favor the GMM if $\Delta {\rm IC} < 0$ and strongly favor the GMM if $\Delta {\rm IC} < -5$. 
Conversely, a positive $\Delta$ IC favors the SGM while $\Delta {\rm IC} > 5$ strongly favors 
the SGM. 

\begin{deluxetable}{rccc}
\tablecaption{Minimum $\dM$ (in mag) with strong evidence for GMM.}
\tabletypesize{\footnotesize}
\tablewidth{0pt}
\tablecolumns{12}
\tablehead{ \colhead{$N$}  & \colhead{AIC} &  \colhead{BIC} &\colhead{DIC}  }
\startdata
100  & 0.40   &  0.45  & 0.41   \\
1000  & 0.21  &  0.25  & 0.23   \\
2500  & 0.12  &  0.21  & 0.23  \\
10000  & 0.10  &  0.14  & 0.10   
\enddata
\label{table:IC}
\end{deluxetable}

\begin{figure}
	\plotone{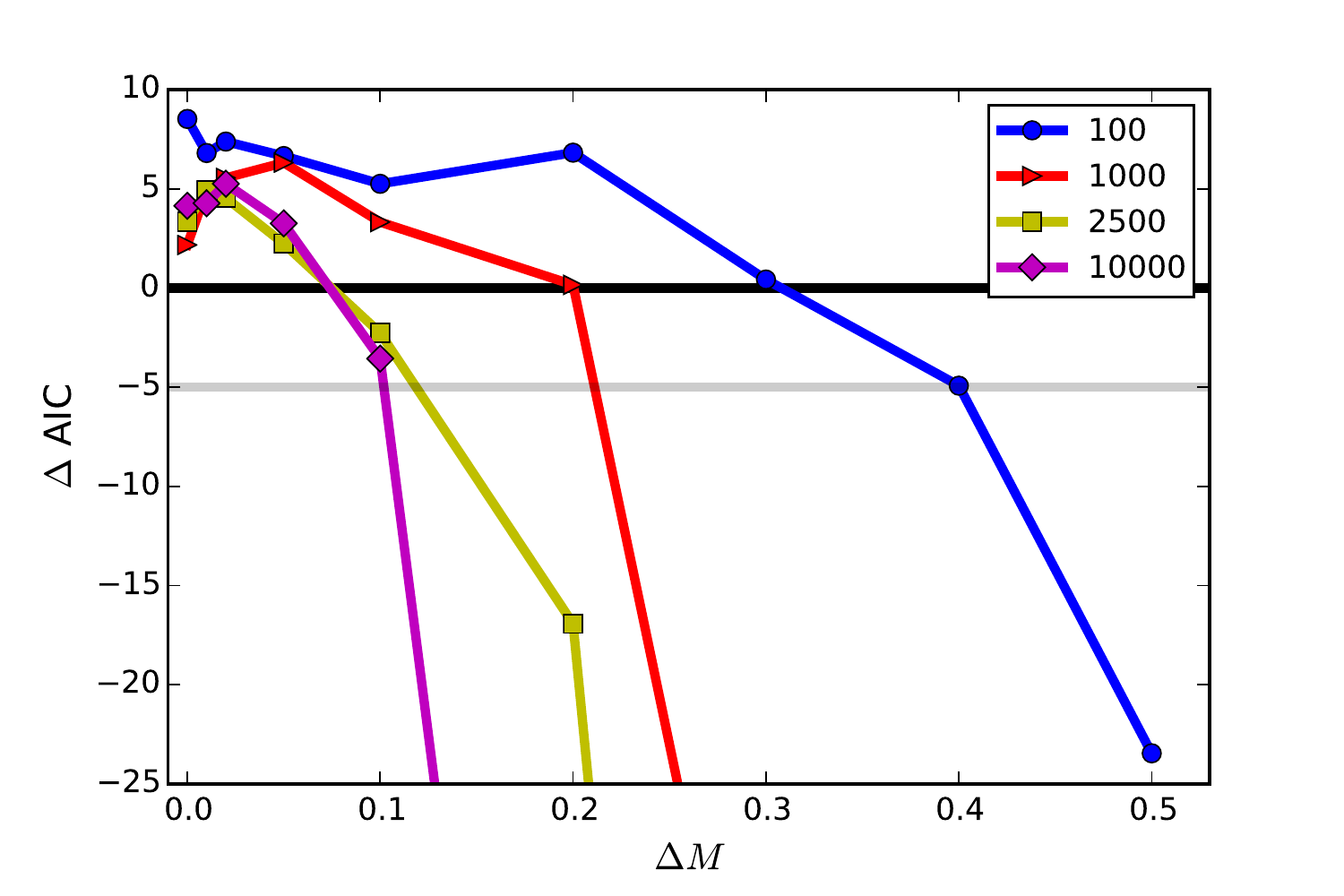}
	\caption{AIC(GMM)$-$AIC(SGM) for N=[100,1000,2500,1000] as a function of the separation of peaks.  GMM is considered strongly favored once $\Delta$AIC$<-5$. 
}
	\label{fig:IC}
\end{figure}

We look for the minimum $\dM$ for each $N$ that strongly favors the GMM. 
The results of this comparison are summarized Table~\ref{table:IC} for all of the IC and in 
Fig.~\ref{fig:IC} for the AIC alone. The AIC, BIC, and DIC all give very comparable results. 
Notice that $\dM$ must be relatively large in order for the 
IC to indicate that the data demand a two-population model of SNeIa. Indeed, a data set 
of $N \gtrsim 10,000$ SNeIa is required in order for the IC to prefer strongly the GMM with 
$\dM\sim0.1$ over the SGM.

There is an important point regarding the interpretation of the results of this section 
in conjunction with those of the previous subsections. 
The fact that the data may not {\em demand} 
a GMM to describe SNeIa does {\em not} mean that a multiple-population SNeIa model is 
not {\em necessary}. As we have shown, statistically significant biases in cosmological parameters 
can be inferred when two-population data are analyzed as a single population, 
even when the information criteria do not unambiguously demand the GMM rather than the SGM. 
If by ``necessary" one means that the model is needed in order to infer unbiased cosmological 
parameters, then the GMM may be necessary even when the IC yield only marginal evidence. 
IC that do not clearly demand the more complex model (the GMM in this case) are not sufficient 
justification for using only the simpler model (the SGM in this case) in cosmological analyses because 
significant parameter biases may still be realized using the simpler approach.


\section{Discussion}\label{sec:discussion}

\subsection{Usage of the SGM}

The SGM was meant to be representative of the latest supernova cosmology analysis, namely the Joint Lightcurve Analysis (JLA); however, the SGM cannot be directly compared to the JLA.
Unlike the SGM, the JLA further standardizes each supernova by applying an offset to the absolute magnitude of each supernova using an empirically-derived step function in host galaxy mass. 
This standardization follows from the observed Hubble residual trend with host galaxy properties that was one of the motivations for introducing multiple populations. 
Leaving out the host galaxy standardization enables this present study to avoid any unintended bias from using the step function, conceptually compare the SGM to the GMM, and create a general framework that can be applied to other systematics.
The goal of this paper is not to implement a new model for the correlation between the SN Hubble residual and host-galaxy properties, but to introduce a statistical framework in which to implement a future model.

The likelihoods for the SGM and JLA are the same except that the JLA utilizes the host galaxy mass standardization and a full covariance matrix. 
JLA uses a $\chi^2$ minimization for parameter estimation, which is equivalent to maximizing a Gaussian likelihood. 
The JLA uses a frequentist approach with $\chi^2$ minimization, but we use the SGM to explore parameter space through Bayesian statistics with MCMC. 
However, $\chi^2$ minimization and the SGM likelihood analysis both use a Gaussian single-point estimate of the SN corrected brightness to infer cosmological parameters. 
Using single-point estimate $(\mu, \sigma)$ does not provide framework to deal with insufficient population modeling and data with large error bars on parameters used for systematics modeling\footnote{The mass of each host galaxy is determined from photometry in the JLA sample has a typical uncertainty of $\sim0.8$ dex.}, both of which are found in the current data sets.
 This present paper shows that updating the likelihood to incorporate non-Gaussian effects can remove bias on cosmology without precise modeling of the underlying populations.
 
\subsection{Connection To Astrophysical Properties}

A relationship with host galaxy mass is currently used to correct SN Ia apparent brightness; however, 
host mass must be an indicator of a different galactic property that has a connection to the brightness of a supernova 
such as local metallicity, star formation rate, and stellar population age \citep{Johansson13}. 
One possible explanation for the host mass effect is different progenitor ages.
The overall mass of the galaxy is correlated with progenitor age through stellar population ages. 
SNeIa occur in both active and passive star forming regions, which implies that they have both short delay times ($\sim 100-500$ Myr) and long delay times ($\sim 5$ Gyr) 
between progenitor formation and supernova event
\citep{Mannucci05, Scannapieco05, Mannucci06, Sullivan06}. 
The different progenitor ages could be motivated by different channels for a thermonuclear explosion: single degenerate (SD) where a white dwarf accretes matter from a main sequence or red giant companion \citep{WhelanIben}
and double degenerate (DD) where two white dwarfs merge \citep{Webbink84}. 
The SD and DD can both explain the population with short delay times; however, SD models do not support the long delay times (\cite{Greggio05}, see \cite{Maoz14} for comprehensive review). 

Several papers have begun to examine the connections between host galaxy mass and stellar population ages. 
\cite{Johansson13} showed that the stretch-host galaxy mass relationship is caused by the correlation between host galaxy mass and stellar population age. 
\cite{Childress13a} fit the Hubble residual versus host galaxy mass with different functional forms and examined different physical causes of the relationship. 
They found the best physical link to the step function was the evolution of the prompt fraction of SN Ia progenitors, but the fit is not adequate enough to be the only source of the effect. 
\cite{Childress14} focused on modeling stellar population age as a function of host galaxy mass at different redshifts.
The paper showed a bimodal distribution in progenitor age versus stellar mass and that this bimodality is evident out to a redshift  $\la 0.5$.
These results clearly motivate adopting a GMM approach where the two populations changing with redshift.
Unfortunately, the way the populations evolve with redshift is 
determined through star formation histories and delay time distributions, which is considerably more complicated than the simple linear evolution probed here. 
Creating better astrophysical models for the evolution of systematics is an active area of research, 
and we present this generalized PDF approach as the appropriate framework to incorporate them into.


 \section{Conclusion}\label{sec:conclusion}
This paper explored expanding supernovae analyses into a broader scope with a generalized likelihood model.  
For illustration we used a toy example of two-population GMM with a simple linear evolution in relative population with redshift.
We explored different distributions of likelihood functions and
showed that in mock data sets using our toy GMM example
multiple SNeIa sub-populations may lead to significant biases in cosmological parameters inferred from SNeIa data. 
In particular, when $N=1,000$ and $\dM=0.1$~mag, biases may be 2-4 times that of the statistical uncertainty. 
Incorporate this model into the PDF removes systematic errors (biases) in inferred 
cosmological parameters at a small statistical cost, roughly 2\% in the marginalized uncertainty on $w$.
Large data sets (N $>$ 10,000) are necessary to yield unambiguous evidence of multiple populations according to various 
model selection criteria. However, even when model selection does not clearly favor multiple populations, the presence of multiple 
populations in the data can severely bias cosmological parameters. 
Our approach of modeling the possibility of multiple populations not only mitigates biases from them, but also 
yields a small penalty in precision if there is only one population.

The existence of multiple populations is still being debated as seen in \cite{Jones15}, which advocates for a single population; however,
a GMM likelihood has the capability of determining if there is only one population and thus is a more rigorous way to analyze the data to ensure more systematics are included. 

We have assumed an example model of two populations with a difference in the absolute magnitude, but there are clearly other channels in which separate populations might be expressed depending on the astrophysical cause. 
It is possible that a different supernova property can better parameterize the stellar population age of the progenitor.
If we did not use the width-color-corrected apparent magnitude, then the apparent magnitudes would be defined as $M_X \equiv M_{ {\rm Bband}, X} - \alpha x_1 + \beta  \mathcal{C}$,
where $x_1$ is the stretch calculated from each supernova light curve, 
$\alpha$ is the stretch parameter determined for the entire supernova population, 
$\mathcal{C}$ is the color of each supernova at time of maximum light, 
and $\beta$ is the color parameter determined for the entire supernova population.
One example has been provided by \cite{Milne14} which shows two different populations with a difference in near ultraviolet (NUV) $u-v$ color of $0.4$~magnitudes ($0.1$~mag in $b-v$) with the relative fractions of populations evolving with redshift. 
This color dependence would fit nicely into our framework since we could alternatively model the absolute magnitude as $M_X \equiv M_{\rm Bband} - \alpha x_1 + \beta_X  \mathcal{C}$. 

This framework is tested with the host galaxy mass dependence as an example; however, it is suitable for accounting for any systematic that may have multiple values based on supernova parameters. 
For example,
surveys with different selection effects could also be included as different PDFs, either in intrinsic distribution or in redshift evolution, for each survey.
Corrections for Malmquist bias \citep{Malmquist36} could be handled more cleanly by using the full PDF instead of using the mean computed correction for the sample \citep[e.g.,][]{Perrett10,Conley11,Rest14,Scolnic14,Betoule14} or priors on the light-curve fitting parameters applied on per-object basis \citep[e.g.,][]{Wood-Vasey07}.
Currently forward-modeling approaches that simulate entire surveys \citep[e.g., SNANA][]{Kessler09,Kessler10} carry through this modeling all the way; 
we believe there can be significant gains in translating much of this information into empirical PDFs that can then be interpolated and used in a generalized full-likelihood fitting 
(work towards this has begun in \cite{Rubin15}).

Supernova cosmology would benefit from incorporating a non-Gaussian likelihood with an MCMC analysis to model the many 
systematics involved in order to remove biases with a minimal precision loss. 
\

\acknowledgments
We thank the referee for constructive comments, which have improved this paper.
We thank Phil Marshall and Carles Badenes for useful discussions.
KAP was supported in this work by NSF AST-1028162, and by the U.S. Department of Energy, Office of Science, Office of Workforce Development for Teachers and Scientists, Office of Science Graduate Student Research (SCGSR) program. 
The SCGSR program is administered by the Oak Ridge Institute for Science and Education for the DOE under contract number DE-AC05-06OR23100. 
M.W.-V. was supported in part by DOE DE-SC0011834 and through sabbatical support from Stanford/SLAC.
The work of ARZ is supported in part by the DOE through grant DE-SC0007914 and by the Pittsburgh 
Particle physics, Astrophysics, and Cosmology Center (Pitt PACC) at the University of Pittsburgh.

\bibliographystyle{apj}
\bibliography{Ponder_GMM}

\end{document}